\def\aprge{\buildrel > \over {_{\sim}}}

\newcommand{\cm}{\rm \,cm}
\newcommand{\mm}{\rm \,mm}
\newcommand{\m}{\rm \,m}

\newcommand{\km}{\rm \,km}

\newcommand{\s}{\rm \,s}

\newcommand{\sr}{\rm \,sr}

\newcommand{\mg}{\rm \,mg}

\newcommand{\g}{\rm \,g}
\newcommand{\hg}{\rm \,hg}
\newcommand{\MeV}{\rm \,MeV}
\newcommand{\eV}{\rm \,eV}

\newcommand{\GeV}{\rm \,GeV}
\newcommand{\TeV}{\rm \,TeV}

\newcommand{\MHz}{\rm \,MHz}

\newcommand{\anue}{$\bar{\nu}_e$}

\newcommand{\lsim}{\lower .5ex\hbox{$\buildrel < \over {\sim}$}}
\newcommand{\gsim}{\lower .5ex\hbox{$\buildrel > \over {\sim}$}}

\newcommand{\stitolo}{
Status Report of the MACRO Experiment for the year 2001

}
\newcommand{\scollabor}{

MACRO Collaboration                    

}
\newcommand{\sautori}{

  M.~Ambrosio$^{\mbox{12}}$, R.~Antolini$^{\mbox{7}}$, G.~Auriemma$^{\mbox{14,a}}$,
D.~Bakari$^{\mbox{2,17}}$, A.~Baldini$^{\mbox{13}}$,  
  G.~C.~Barbarino$^{\mbox{12}}$, B.~C.~Barish$^{\mbox{4}}$,  G.~Battistoni$^{\mbox{6,b}}$, 
Y.~Becherini$^{\mbox{2}}$, R.~Bellotti$^{\mbox{1}}$
% ...

}
\newcommand{\sfirmatoda}{

\collabor                            

}
\newcommand{\sistituzioni}{
                                      
$^{\mbox{1}}$                         
Dipartimento di Fisica dell'Universit\`{a} di Bari and INFN, 70126 Bari, Italy 
\\ 
$^{\mbox{2}}$                         
Dipartimento di Fisica dell'Universit\`{a} di Bologna and INFN, 40126 Bologna, Italy \\
% ...
$^{\mbox{4}}$                         
California Institute of Technology, Pasadena, CA 91125, USA\\
% ...
$^{\mbox{6}}$                         
Laboratori Nazionali di Frascati dell'INFN, 00044 Frascati (Roma), Italy\\
$^{\mbox{7}}$                         
Laboratori Nazionali del Gran Sasso dell'INFN, 67010 Assergi (L'Aquila), Italy\\
% ...
$^{\mbox{12}}$                         
Dipartimento di Fisica dell'Universit\`{a} di Napoli and INFN, 80125 Napoli, Italy\\
$^{\mbox{13}}$                         
Dipartimento di Fisica dell'Universit\`{a} di Pisa and INFN, 56010 Pisa, Italy\\
$^{\mbox{14}}$                         
Dipartimento di Fisica dell'Universit\`{a} di Roma {}``La
Sapienza\char`\"{} and INFN, 00185 Roma, Italy\\
% ...
$^{\mbox{17}}$                         
L.P.T.P., Faculty of Sciences, University Mohamed I, B.P. 524 Oujda, Morocco\\
% ...
$^{\mbox{a}}$                         
Also Universit\`{a} della Basilicata, 85100 Potenza, Italy\\
$^{\mbox{b}}$                         
Also INFN Milano, 20133 Milano, Italy \\
% ...
}                         

\newcommand{\sabst}{

In this 2001 status report of the MACRO experiment, results are presented
on atmospheric neutrinos and neutrino oscillations, high energy neutrino
astronomy, searches for WIMPs, search for low energy stellar gravitational
collapse neutrinos, stringent upper limits on GUT magnetic monopoles,
nuclearites and lightly ionizing particles, high energy downgoing
muons, primary cosmic ray composition and shadowing of primary cosmic
rays by the Moon and the Sun.

}
\documentclass[12pt,twoside]{article}
\usepackage{epsfig}                     
\newcommand{\preprintnum}{
\begin{flushright}
\begin{tabular}{c}
{\bf LNGS/exp-01/02} \\ \hline
{\bf 5 Jun 02} \\ 
\end{tabular}
\end{flushright}}
\newcommand{\titolo}{
\begin{center}\Large{\bf{
\stitolo
}}\end{center}}
\newcommand{\collabor}{
\begin{center}
\Large{
\scollabor                             
}    
\end{center}}
\newcommand{\autori}{
\begin{center}
\sautori
\end{center}}
\newcommand{\istituzioni}{
\begin{center}
\small{\it{
\sistituzioni
}}                       
\end{center}}                         
\newcommand{\abst}{
\begin{abstract}
\sabst               
\end{abstract}}
\newcommand{\firmatoda}{
\begin{center}
\sfirmatoda
\end{center}}
\begin{document}
\pagestyle{empty}
\enlargethispage{2cm}
\preprintnum
\titolo
\firmatoda
\autori
\istituzioni
\abst
\clearpage     
\setcounter{page}{2} 
\pagestyle{plain}                
%

%
\section{Introduction}

MACRO was a large area multipurpose underground detector designed
to search for rare events in the cosmic radiation. It was optimized
to look for the supermassive magnetic monopoles predicted by Grand
Unified Theories (GUT) of the electroweak and strong interactions;
it could also perform measurements in areas of astrophysics, nuclear,
particle and cosmic ray physics. These include the study of atmospheric
neutrinos and neutrino oscillations, high energy \( (E_{\nu }\, \, \gsim \, \, 1\, \, \GeV ) \)
neutrino astronomy, indirect searches for WIMPs, search for low energy
(\( E_{\nu }\, \, \gsim \, \, 7\, \,  \)MeV) stellar collapse neutrinos,
studies of various aspects of the high energy underground muon flux
(which is an indirect tool to study the primary cosmic ray composition,
origin and interactions), searches for fractionally charged particles
and other rare particles that may exist in the cosmic radiation. 

The mean rock depth of the overburden is \( \simeq 3700 \) m.w.e.,
while the minimum is \( 3150 \) m.w.e. This defines the minimum muon
energy at the surface at \( \sim 1.3{\TeV } \) in order to reach
MACRO. The average residual energy and the muon flux at the MACRO
depth are \( \sim 320{\GeV } \) and \( \sim 1{\m }^{-2}{\textrm{h}}^{-1} \),
respectively. 

The detector was built and equipped with electronics during the years
\( 1988-1995 \). It started data taking with part of the apparatus in \( 1989 \); it was completed
in \( 1995 \) and it was running in its final configuration until
December 19, 2000. It may be worth pointing out that all the physics and
astrophysics items proposed in the 1984 Proposal were covered and good
results were obtained on each of them, even beyond the most rosy anticipations.

The highlights of the new results have been presented at the 2001
summer conferences (in particular at the Int. Cosmic Ray Conf. (ICRC)
in Hamburg, at the 2001 European HEP in Budapest, at TAUP 2001 at
Gran Sasso and at the NATO Advanced Research Workshop in Oujda, Morocco).
One of the main results is the evidence for anomalies in the atmospheric
\( \nu _{\mu } \) flux, which are well interpreted in terms of \( \nu _{\mu }\rightarrow \nu _{\tau } \)
oscillations.

We shall give a short summary of the detector and of its performances;
this will be followed by an overview of the main physics and astrophysics
results obtained by MACRO. A complete list of MACRO papers is given in
\cite{mac1}-\cite{mac36}; other information may be found in http://www.df.unibo.it/macro/pub1.htm.

In the year \( 2001 \) four papers were published on refereed journals;
they concerned high energy neutrino astronomy with the MACRO detector
\cite{mac33}, the preference for \( \nu _{\mu }\rightarrow \nu _{\tau } \)
oscillations over \( \nu _{\mu }\rightarrow \nu _{s} \) \cite{mac34},
a technical paper on the MACRO detector \cite{mac35} and a combined
analysis for a search for magnetic monopoles \cite{mac36}. Several
results appeared in preliminary form in 14 paper contributions that
were published in various physics conference proceedings \cite{mac37}-\cite{mac51}.
They concerned the study of high and low energy atmospheric neutrinos,
the use of multiple Coulomb scattering for determining neutrino energies,
high energy muon neutrino astronomy, several rare particle searches,
the observation of the moon and sun shadow of high energy primary
cosmic rays and several aspects of {}``muon astronomy{}''.

\section{The Detector}

The MACRO detector had a modular structure: it was divided into six
sections referred to as supermodules. Each active part of one supermodule
had a size of \( 12.6\times12 \times9 .3{\m }^{3} \) and had a separate
mechanical structure and electronics readout. The full detector had
global dimensions of \( 76.5\times12 \times9 .3{\m }^{3} \) and provided
a total acceptance to an isotropic flux of particles of \( \sim 10,000{\m }^{2}{\sr }. \)
The total mass was \( \simeq 5300\, \,  \)t.

Redundancy and complementarity have been the primary goals in designing
the experiment. Since no more than few magnetic monopoles could be
expected, multiple signatures and ability to perform cross checks
among various parts of the apparatus were important.

\begin{figure}
  \begin{center}
  \mbox{ 
\epsfysize=8cm
         \epsffile{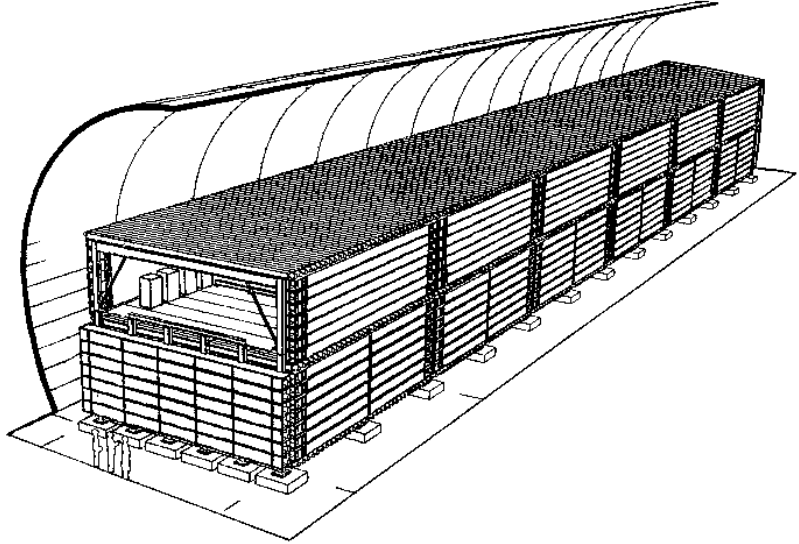} 
}   
\caption{\label{fig1}\small General layout of the MACRO detector which was installed in Hall B of the LNGS.
Overall dimensions of the active part were $76.5\times12\times9.3~{\m}^3.$  \cite{mac35}
}
\end{center}
\end{figure}

\begin{figure}
\vspace{-1cm}
  \begin{center}
  \mbox{ 
\epsfysize=8cm
         \epsffile{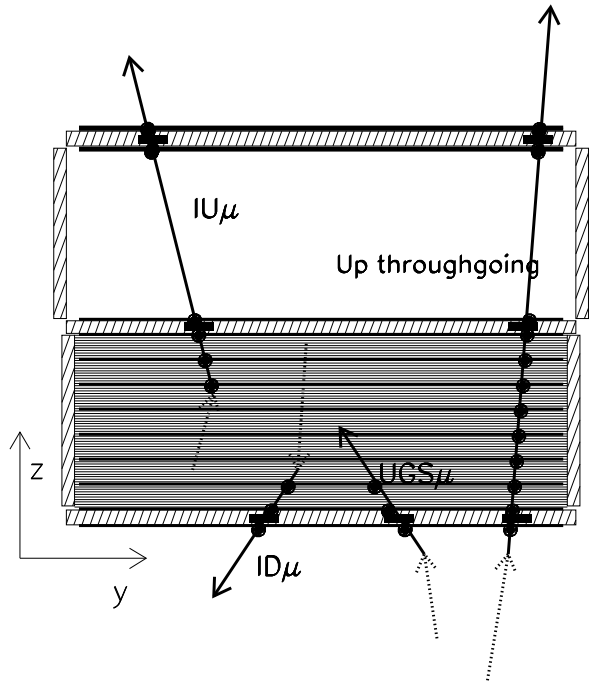} 
}   
\caption{\label{fig2}\small Vertical cross section of the detector and sketch of different event
topologies induced by $\nu_\mu $ interactions in or around MACRO.
The black points and the black rectangles represent streamer tubes and
scintillator hits, respectively. Tracking was performed by the streamer
tubes;  the time-of-flight of the muons was measured by the
scintillators for {\it Up Semicontained } (Internal upgoing - IU $\mu$)
and {\it Upthroughgoing }  events (and also for downgoing muons).}
\end{center}
\end{figure}

The detector was composed of three sub-detectors: liquid scintillation
counters, limited streamer tubes and nuclear track detectors. Each
one of them could be used in \lq\lq stand-alone\rq\rq~and in \lq\lq combined\rq\rq~mode.
A general layout of the experiment is shown in Fig. \ref{fig1}. Notice
the division in the \textit{lower} MACRO and in the  \textit{upper} part, often
referred to as the \textit{Attico}; the inner part of the \textit{Attico} was
empty and lodged the electronics. The mass of the \textit{lower}
MACRO was \( \simeq 4200\, \,  \)t, mainly in the form of boxes filled with
crushed
Gran Sasso rock. Fig. \ref{fig2} shows a cross section of the apparatus.

{\bf The scintillation subdetector.} Each supermodule contained 77
scintillation  counters, divided into three horizontal planes (bottom,
center, and top) and two vertical planes (east and west). In the lower part,
the bottom and center horizontal planes had 16 scintillation counters, the
east and west vertical planes had 7 counters each. In the \textit{Attico},
the top plane had 17 scintillation counters, the east and west vertical
planes had 7 counters each. The lower part of the north and
south faces of the detector were covered by vertical walls with seven
scintillation counters each. The upper parts of these faces were left
open in order to allow access to the readout electronics. 

The active volume of each horizontal scintillation counter was \( 11.2\times 0.73\times 0.19{\m }^{3} \),
while for the vertical ones it was  \( 11.1\times 0.22\times 0.46{\m }^{3} \).
All scintillator boxes were filled with a mixture of high purity mineral oil (\( 96.4\, \% \)) and
pseudocumene (\( 3.6\, \% \)), with an additional \( 1.44{\g } \)/l
of PPO and \( 1.44{\mg } \)/l of bis-MSB as wavelength shifters.
The horizontal counters were seen by two \( 8^{\prime \prime } \)
photomultipliers (PMTs) and the vertical counters by one \( 8^{\prime \prime } \)
PMT at each end. Each PMT housing was equipped with a light collecting
mirror. The total number of scintillators was \( 476 \) (294 horizontal
and 182 vertical) with a total active mass of almost \( 600\, {\textrm{tons}} \).
Minimum ionizing muons when crossing vertically the \( 19{\cm } \)
of scintillator in a counter release an average energy of \( \simeq 34{\MeV }
\) and were measured with a timing and longitudinal position resolution
of \( \simeq 500\, {\textrm{ps}} \) and \( \simeq 10{\cm } \), respectively.

The scintillation counters were equipped with specific triggers for
rare particles, muons and low energy neutrinos from stellar gravitational
collapses. The Slow Monopole Trigger (SMT) was sensitive to magnetic
monopoles with velocities from about \( 10^{-4}c \) to \( 10^{-2}c \),
the Fast Monopole Trigger (FMT) was sensitive to monopoles with velocities
from about \( 5\times10 ^{-3}c \) to \( 5\times10 ^{-2}c \), the
Lightly Ionizing Particle trigger was sensitive to fractionally
charged particles, the Energy Reconstruction Processor (ERP) and {}``CSPAM{}''
were primarily muon triggers (but used also for relativistic monopoles)
and the gravitational collapse neutrino triggers (the Pulse Height
Recorder and Synchronous Encoder --PHRASE-- and the ERP) were optimized
to trigger on bursts of low energy events in the liquid scintillator.
The scintillator system was complemented by a \( 200{\MHz } \) waveform
digitizing (WFD) system used in rare particle searches, and in any occasion 
where knowledge of the PMT waveform was useful.

{\bf The streamer tube subsystem.} \textit{The lower part} of the
detector contained ten horizontal planes of limited streamer tubes,
the middle eight of which were interleaved by seven rock absorbers
(total thickness \( \simeq 360{\g }{\cm }^{-2} \)). This sets a \( \simeq 1{\GeV } \)
energy threshold for muons vertically crossing the lower part of the
detector. At the top of the \textit{Attico} there were four horizontal streamer
tube planes, two above and two below the top scintillator layer. On
each lateral wall six streamer tube planes sandwiched the corresponding
vertical scintillator plane (three streamer planes on each side).
Each tube had a \( 3\times 3{\cm }^{2} \) cross section and was 
\( 12{\m } \) long. The total number of tubes was \( 50304 \),
all filled with a gas mixture of \( He \) (\( 73\, \% \)) and n-pentane
(\( 27\, \% \)). They were equipped with \( 100{\mu } \) \textit{Cu/Be}
wires and stereo pickup strips at an angle of \( 26.5^{\circ } \).
The tracking resolution of the streamer tube
system was \( \simeq 1{\cm } \), corresponding to an angular accuracy of 
\( \simeq 0.2^{\circ } \) over the 9.3 m height of MACRO. The real angular 
resolution was limited to \( \simeq 1^{\circ } \) by the multiple Coulomb 
scattering of muons in the rock above the detector. The streamer
tubes were read by \( 8 \)-channel cards (one channel for each wire)
which discriminated the signals and sent the analog information (time
development and total charge) to an ADC/TDC system (the QTP). The
signals were used to form two different chains (Fast and Slow) of
TTL pulses, which were the inputs for the streamer tube Fast and Slow
Particle Triggers. In the 11 years of operation only 50 wires were lost.

{\bf The nuclear track subdetector} was deployed in three planes,
horizontally in the center of the lower section and vertically on
the East and North faces. The detector was divided in \( 18126 \)
modules, which could be individually extracted and substituted. 
Each module (\( \sim 24.5\times 24.5\times 0.65 {\cm }^{3} \) ) was 
composed of three layers of CR\( 39 \), three layers
of Lexan and \( 1\, {\textrm{mm}} \) Aluminium absorber to stop nuclear
fragments.

{\bf The Transition Radiation Detector (TRD).} A TRD was installed in part of the \textit{Attico}, right
above the central horizontal scintillator plane of the main detector.
It was composed of three individual modules (overall dimensions \( 6\times 6\times 2{\m }^{3} \))
and it was made of \( 10{\cm } \) thick polyethylene foam radiators
and proportional counters; each counter measured \( 6\times 6\times 600{\cm }^{3} \)
and was filled with \( Ar \) (\( 90\, \% \)) and \( CO_{2} \) (\( 10\, \% \)).
The TRD provided a measurement of the muon energy in the range of
\( 100{\GeV }<E<930{\GeV } \); muons of higher energies could also
be detected and counted.

Fig. \ref{fig3} shows four photographs of the Hall B taken from its
south side: (a) 1987: before starting construction; (b) 1990: the 1st
lower supermodule was taking data, while the second and the third were
under construction; (c) the full MACRO detector in 1995 (a safety
stairs and a ventilation system were added later in front of the
apparatus); (d) Hall B empty again in 2001. 
(You can find the pictures at this address:\\
http://www.df.unibo.it/margiotta/rep\_01/
with the names: fig3a.eps, fig3b.eps, fig3c.eps, fig3d.eps.)
\begin{figure}
 \vspace{-2.cm}
  \begin{center}
  \mbox{ 
\vspace{5cm.}
\epsfysize=7cm
         \epsfxsize=7cm
  \epsfysize=7cm
  \epsfxsize=7cm
}   
 \vspace{5.5cm}
{\small \hskip 1.0 truecm (a) \hskip 7.5 truecm (b) \vspace{5.5cm}}
  \mbox{\vspace{.5cm}
        \epsfysize=7cm
  \epsfysize=7cm
  \epsfxsize=7cm
}   
\vspace{.5cm}
{\small \hskip 6.0 truecm (c) \hskip 7.5 truecm (d)}

 \end{center}
  \caption {\label{fig3}\small Photographs of Hall B taken from its south
         side: (a) In 1987 just before starting construction; (b) in 1990
         when the first lower supermodule was taking data while the second
         and the third were under construction; (c) in 1995 when the
         completed MACRO detector started data taking (notice that safety
         stairs and a ventilation system were added later in front of the
         apparatus; (d) Hall B empty in 2001. (You can find the pictures at this address:
 http://www.df.unibo.it/margiotta/rep\_01/
with the names: fig3a.eps, fig3b.eps, fig3c.eps, fig3d.eps.)}
\end{figure}

\begin{figure}
 \begin{center}
 \hspace{1.cm}
  \mbox{ \epsfysize=3cm
         \epsffile{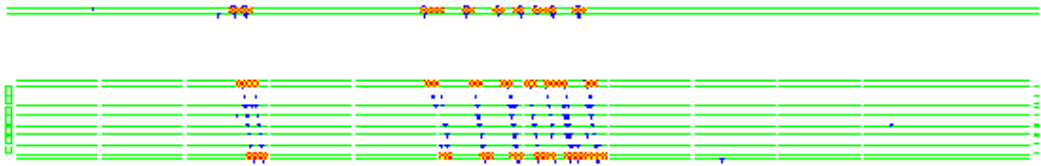} }
 \end{center}
  \vspace{-.5cm}
\caption {\label{fig4}\small MACRO Event Display. A group of 11 downgoing
  muons as observed by part of the lateral view.}
\end{figure}

Fig. \ref{fig4} shows a {}``group{}'' of 11 downgoing muons as
seen  in the lateral view (wire view) 
by the MACRO Event Display (which also included a strip view and side views).

\section{Atmospheric  neutrino oscillations}
\begin{figure}
 \vspace{-2.cm}
 \begin{center}
  \mbox{ \epsfysize=10.5cm
         \epsffile{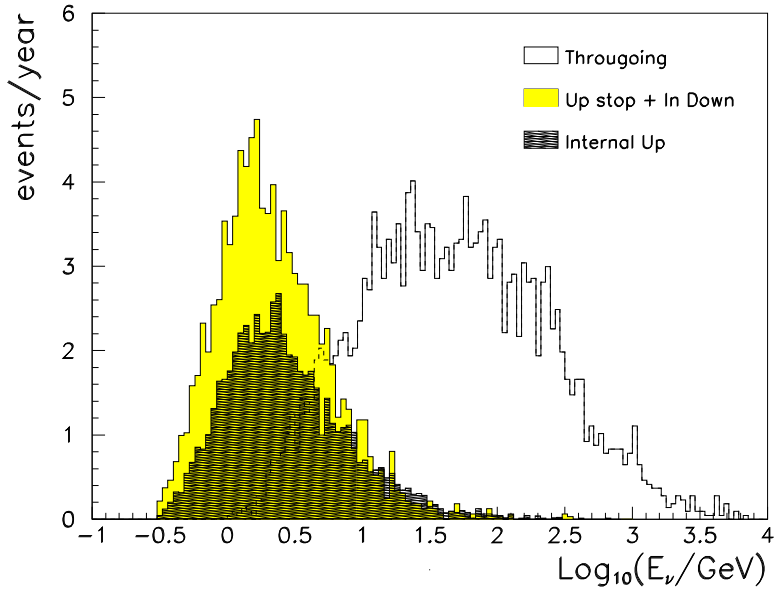} }
 \end{center}
  \vspace{-1.cm}
\caption {\label{fig5}\small Distributions of the parent muon neutrino
energies giving rise to the different event topologies, upthroughgoing,
upsemicontained and upstopping plus downsemicontained, with median neutrino 
energies of approximately 50, 4.2 and 3.5 GeV, respectively.}
\end{figure}

Upward going muons are identified using the streamer tube system (for
tracking) and the scintillator system (for time-of-flight measurement).
A rejection factor of at least \( 10^{7} \) is needed, and was reached,  in order to
separate upgoing muons from the  background due to the
downgoing muons. Fig. \ref{fig2} shows a sketch of the  different  neutrino event
topologies  analyzed: Upthroughgoing muons,
Upsemicontained (also called Internal Upgoing muons, IU),  Upgoing Stopping
muons (UGS), Internal Downgoing muons (ID).
Fig. \ref{fig5} shows the parent $\nu_\mu$ energy spectra for
the three event topologies, computed by Monte Carlo methods. The
number of events measured and expected for the three topologies are
given in Table \ref{tab:macro}. All the data samples deviate from the MC
expectations; the deviations point out to the same ${\nu_\mu \rightarrow
  \nu_\tau}$ oscillation scenario.

The background on upgoing muons arising from downgoing muons interacting
in the rock around MACRO and giving an upward going charged particle
was studied in detail for upthroughgoing muons in \cite{mac24}. The
selection cuts reduce this background to $< 1 \%$.

\subsection{Upthroughgoing muons}

\begin{figure}[t]
 \vspace{-1.cm}
 \begin{center}
  \mbox{ \epsfysize=8.5cm
         \epsffile{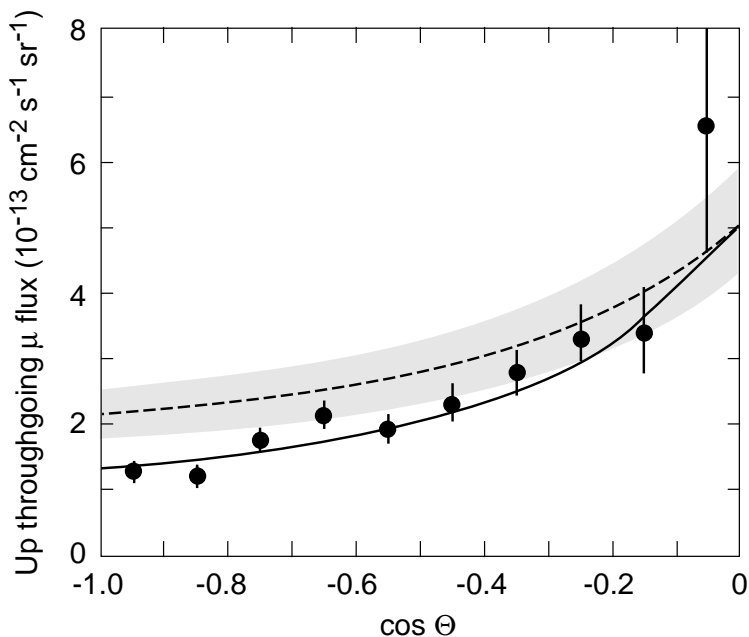}   }
 \end{center}
\vspace{-.3cm}
\caption{\label{fig6}\small Zenith angle distribution of upthroughgoing muons (black points).
The dashed line is the expectation for no oscillations (with a 17 \% scale uncertainty band). The solid line is the fit
for an oscillated muon flux with  maximum mixing and $\Delta m^{2} = 2.5
\cdot 10^{-3}$ eV$^{2}$.}
\end{figure}

The \textit{upthroughgoing muons} come from \( \nu _{\mu } \) interactions
in the rock below the detector; the \( \nu _{\mu } \)'s have a median
energy \( \overline{E}_{\nu }\sim \, \, 50\, \, \GeV  \). The upthroughgoing
muons with \( E_{\mu }>1\GeV  \) cross the whole detector. The time
information provided by the scintillation counters allows the determination
of the direction (versus) by the time-of-flight (T.o.F.) method. The data 
of Fig. \ref{fig6} refer to the running period 3/1989 - 4/1994 with the detector under
construction, and with the full detector till 12/2000; the total livetime
was  6.16 years (full detector equivalent)  \cite{mac17, mac25,  
 mac39,  mac46}. The data deviate in absolute value and in shape from the
MC predictions. This was first pointed at TAUP 1993 and in \cite{mac17} in 1995.

We studied a large number of possible systematic effects that could
affect our measurements:  no significant systematic
problems exist in the detector or in the data analyses. One of the
most significant checks was performed using only the scintillator
system with the PHRASE Wave Form Digitizers, completely independent
of the ERP system.

The measured data have been compared with Monte Carlo simulations.
For the upthroughgoing muon simulation, the neutrino flux computed
by the Bartol group is used \cite{bartol}. The cross sections for the neutrino interactions
were calculated using the deep inelastic parton distributions of ref. \cite{gluck}
The muon propagation to the detector was done 
using the energy loss calculation in standard rock \cite{lohmann}. The total systematic
uncertainty on the expected muon flux, obtained adding in quadrature the errors
from neutrino flux, cross section and muon propagation, is estimated
to be 17 \%. This  uncertainty is mainly a scale error; the error on the shape of the angular distribution is $\sim 5 \%$. 
Fig. \ref{fig6} shows the zenith angle distribution of the measured flux of upthroughgoing
muons. The Monte Carlo expectation for no oscillations is shown as a dashed line. 

To test the oscillation hypothesis, the independent probabilities for
obtaining the observed number of events and the shape of the angular
distribution have been calculated for various parameter values. The
value of \( \Delta m^{2} \) obtained from the shape of the angular
distribution is equal to the value obtained from the observed reduction
in the number of events. 
For \( \nu _{\mu }\rightarrow \nu _{\tau } \)
oscillations, combining the probabilities from the two independent
tests on the shape of the zenith angle distribution and on the total
number of events, the maximum probability is 66\%; the best parameters
are \( \Delta m^{2}=2.5\cdot 10^{-3}\, \, \eV ^{2} \) and maximal
mixing; the result of the fit is the solid line in Fig. \ref{fig6}.  The probability
for no-oscillations is 0.4 \%. 

\begin{figure}
 \begin{center}
  \mbox{ \epsfysize=6cm
         \epsffile{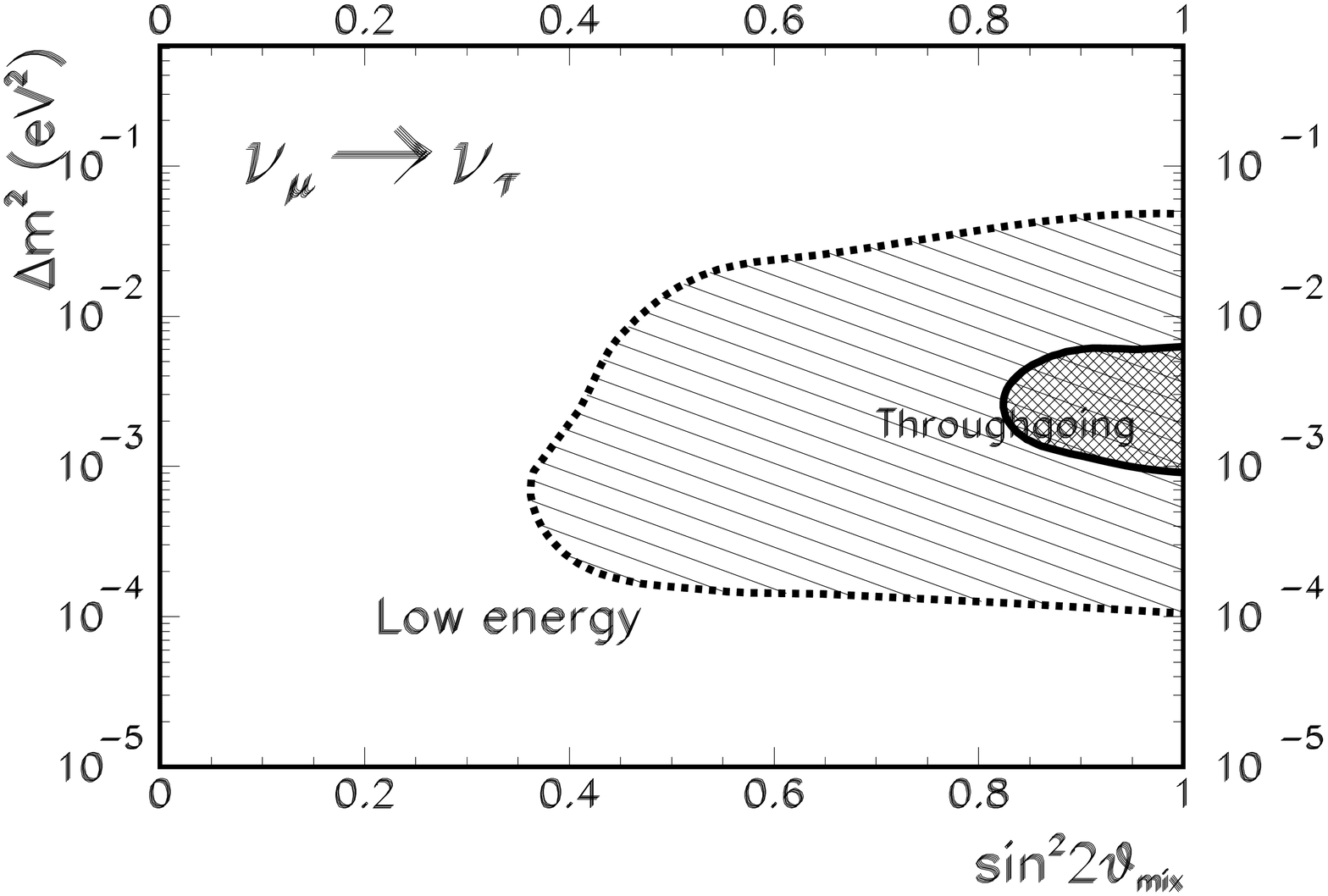}
\epsfysize=5.5cm
         \epsffile{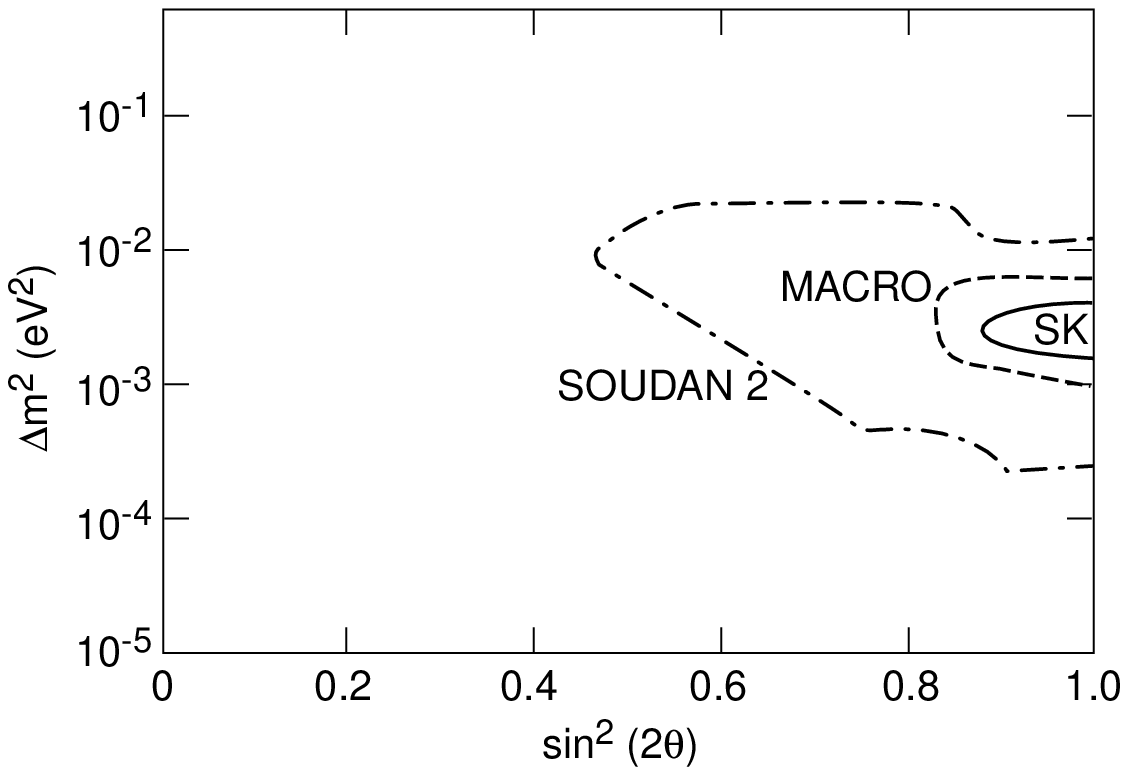}   }
 \end{center}
{\small \hskip 4.5 truecm (a) \hskip 8. truecm (b)}
\caption{\label{fig7}\small  
(a) Allowed regions, at 90 \% c.l., for $\nu_\mu \rightarrow \nu_\tau$
oscillations from the MACRO upthroughgoing muon sample and from the  low energy events.
(b) Comparison with the Soudan 2 \cite{soud2} and SuperK \cite{sk} allowed regions.}
\end{figure}

Fig. \ref{fig7}a shows the allowed regions for the \( \nu _{\mu }\rightarrow \nu _{\tau } \)
oscillation parameters in the \( sin^{2}2\theta -\Delta m^{2}
\)  plane, computed according 
to ref. \cite{feldman} for the upthroughgoing muons and for the low energy events.
The MACRO 90\% c.l. allowed region for \( \nu _{\mu }\rightarrow \nu _{\tau } \)
is compared in Fig. \ref{fig7}b with those obtained by the SuperKamiokande
(SK) \cite{sk} and Soudan 2 experiments \cite{soud2}.

\subsection{Matter effects. $ \nu _\mu \rightarrow \nu _{\tau }$
against $ \nu _{\mu} \rightarrow \nu _{sterile}$ }

Matter effects due to the difference between the weak interaction
effective potential for muon neutrinos with respect to sterile neutrinos
(which have null potential) would produce a different total number
and a different zenith distribution of upthroughgoing muons \cite{mac34}.

\begin{figure}
  \begin{center}
  \mbox{ \epsfysize=8cm
         \epsffile{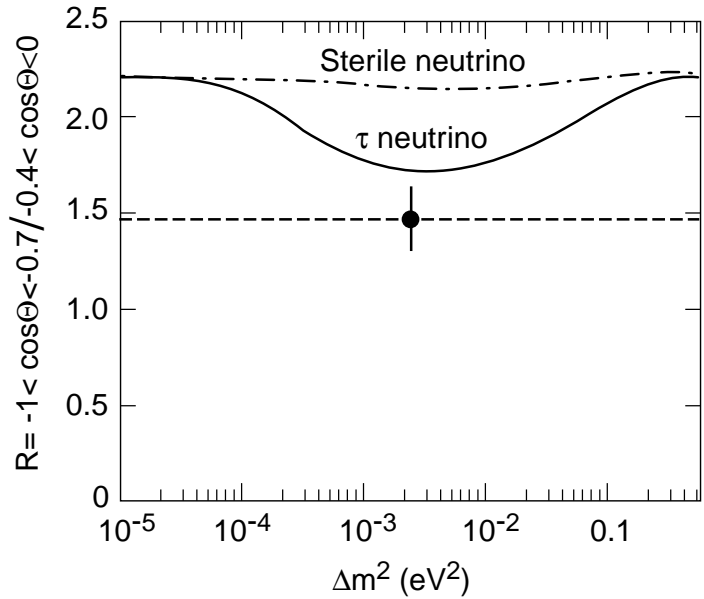} }
\caption{\label{fig8}\small 
Ratio of events with $-1< cos \theta < -0.7$ to events with
$-0.4<cos\theta<0$ as a function of $\Delta m^2$ for maximal mixing. The
black point with error bar is the measured value, the solid line is the
prediction for $ \nu_\mu \rightarrow \nu_\tau $ oscillations, the dash-dotted
line is the prediction for  $ \nu_\mu \rightarrow \nu_{sterile} $ oscillations.
}
\end{center}
\end{figure}

In Fig. \ref{fig8} the measured ratio between the events with 
$ -1 < cos \theta < -0.7$ and the
events with $-0.4 < cos \theta < 0$ is shown as a black point. In this ratio most of the theoretical
uncertainties on neutrino flux and cross section cancel. The remaining
theoretical error is estimated at $\le 5 \%$. The systematic experimental
error on the ratio, due to analysis cuts and detector efficiencies,
is $4.6\%$. Combining the experimental and theoretical errors in quadrature,
a global estimate of $7 \%$ is obtained. MACRO measured 305 events with $ -1 < cos \theta < -0.7$ and 206 with $-0.4 < cos \theta < 0 $; the ratio is R = $1.48 \pm 
0.13_{stat} \pm 0.10_{sys}$. 
For \( \Delta m^{2}=2.5\cdot 10^{-3}\, \, \eV ^{2} \) and maximal
mixing, the minimum expected value of the ratio for \( \nu _{\mu }\rightarrow \nu _{\tau } \)
is $R_{\tau} = 1.72$ while for \( \nu _{\mu }\rightarrow \nu _{s} \) is $R_{sterile} = 2.16$.
The maximum probabilities $P_{best}$  to find a value of  $R_{\tau}$ and of $R_{sterile} $ smaller than the expected ones  are 9.4 \% and 0.06 \% respectively.
Hence the ratio of the maximum probabilities is  $P_{best_\tau} / P_{best_{sterile} } = 157$, so that \( \nu _{\mu }\rightarrow \nu _{s} \)
oscillations are disfavoured at 99\% c.l. compared
to the \( \nu _{\mu }\rightarrow \nu _{\tau } \) channel with maximal
mixing and \( \Delta m^{2}=2.5\cdot 10^{-3}\, \, \eV ^{2} \).

\subsection{$ \ \nu _{\mu }$ energy estimates by multiple
Coulomb scattering of upthroughgoing muons}

The oscillation probability is a function of the ratio $L/E_\nu$. $E_\nu$
may be estimated by measuring the muon energy $E_\mu$, 
which was done using their Multiple Coulomb Scattering (MCS)
in the absorbers.

The r.m.s. of the lateral displacement for a muon crossing the whole
apparatus on the vertical is $\sigma_{MCS}\simeq 10 \cm/ E_\mu (\GeV)$. 
The muon energy $E_\mu$ estimate can be performed up to a saturation
point, occurring when  $\sigma_{MCS}$ is comparable with the detector space resolution.

Two MCS analyses were performed.

The first analysis was made studying the deflection of upthroughgoing
muons with the streamer tubes in digital mode. Using MC methods to
estimate the muon energy from its scattering angle, the data were
divided into 3 subsamples with different average energies, in 2 samples
in zenith angle $\theta$  and finally in 5 subsamples with different average
values of $L/E_\nu$. This method could reach a spatial resolution of $\sim 1 \cm$;
it yielded an  $L/E_\nu$ distribution quite compatible with neutrino oscillations
with the parameters of Section 3.1 \cite{mac52}.

\begin{figure}
\vspace{-2.cm}
  \begin{center}
  \mbox{ \epsfysize=8.5cm
         \epsffile{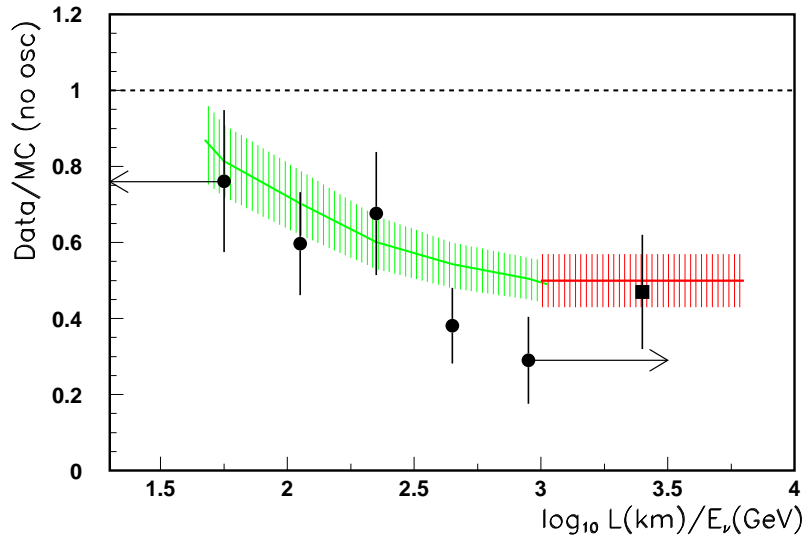} }
\caption{\label{fig9}\small
Ratio (Data/MC) as a function of estimated $L/E_{\nu}$ for the upthrougoing
muon sample (black circles) and the semicontained up-$\mu$ (black square). For
upthroughgoing muons the muon energy was estimated by MCS and $E_{\nu}$ by MC
methods. The shaded regions represent the uncertainties in the MC
predictions assuming $\sin^2 2 \theta=1$ and $\Delta m^2=0.0025$ eV$^2$. 
The horizontal  dashed line at Data/MC=1 is the expectation for no oscillations. 
}
\end{center}
\end{figure}

As the interesting energy region for atmospheric neutrino oscillations
spans from \( \sim 1\, \GeV  \) to some tens of GeV, it is important
to improve the spatial resolution of the detector to push the saturation
point as high as possible. For this purpose, a second analysis was
performed with the streamer tubes in {}``drift mode{}'', using the
TDC's included in the QTP system, originally designed for the search
for magnetic monopoles. To check the electronics and the feasibility
of the analysis, two tests were performed at the CERN
PS-T9 and SPS-X7 beams \cite{mac51}. The space resolution achieved is \( \simeq 3{\mm } \), a factor 3.5
better than in the first analysis. For each muon, seven MCS sensitive
variables were given in input to a Neural Network (NN) previously
trained to estimate the muon energy with MC events of known input energy crossing the detector
at different zenith angles. The NN output allowed to separate the upthroughgoing
muons in 4 subsamples with average energies  of 12, 20, 50 and
102 GeV, respectively. The comparison of their zenith angle distributions
with the predictions of the no oscillations MC shows a  disagreement
at low energies (where there is a deficit of vertical events), while
the agreement is restored at increasing  neutrino energies.
The distribution of the ratio $R = (Data/MC_{no osc})$ obtained by this
analysis is plotted in Fig. \ref{fig9} as a function of
$log_{10}(L/E_\nu)$ \cite{mac41,mac50,mactesi}. The black points with error
bars are the data; the vertical extent of the shaded areas represents
the uncertainties on the MC predictions for \( \nu _{\mu }\rightarrow \nu _{\tau } \)
oscillations with maximal mixing and \( \Delta m^{2}=2.5\cdot 10^{-3}\, \,
\eV ^{2} \). The horizontal dashed line is the expectation without oscillations. 
The last data point (black square) has been obtained from the low
energy IU sample.

\subsection{Low energy data.}

\begin{figure}
 \begin{center}
  \mbox{ \epsfysize=7cm
         \epsffile{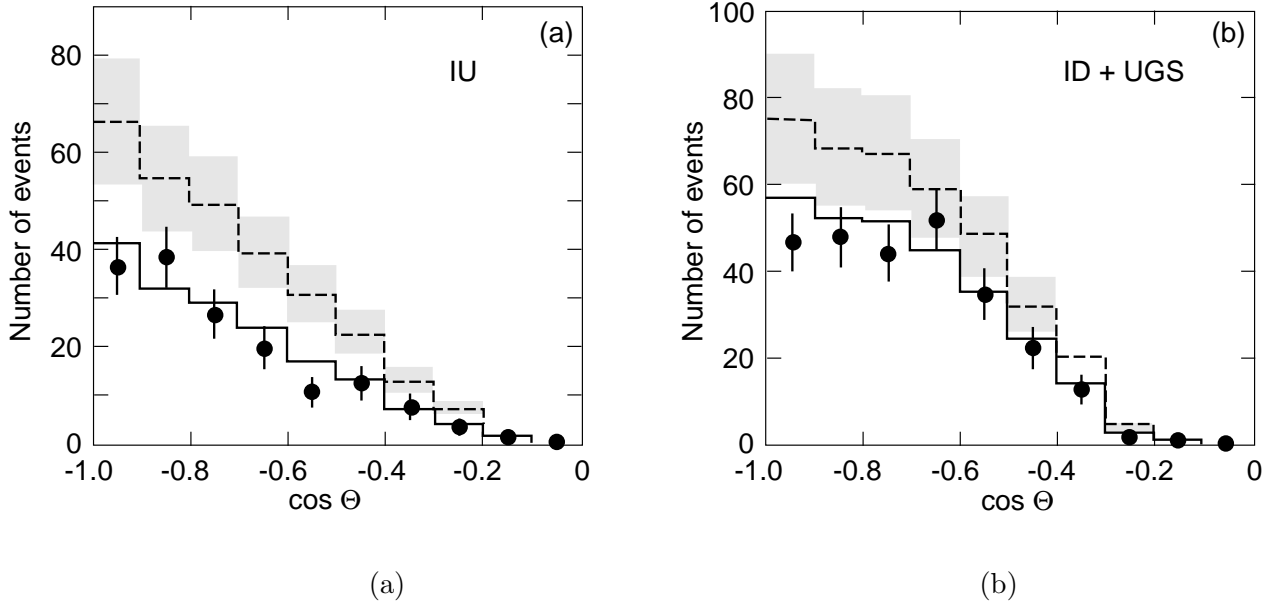}
           }
 \end{center}
{\centering 
{\small \hskip 5.0 truecm (a) \hskip 7.5 truecm (b)}}
\caption{\label{fig10}\small  
Measured zenith distributions (a) for the upsemicontained (IU) events and 
(b) for the upstopping plus the downsemicontained (ID+UGD) events. The black
points are the data, the dashed line at the center of the shaded regions
correspond to MC predictions 
assuming no oscillations. The full line is the expectation
for  $\nu_\mu \rightarrow \nu_\tau$ oscillations with maximal mixing and $\Delta m^2 = 2.5 \cdot 10^{-3 } \eV^2$.}
\end{figure}

The \textit{Internal Upgoing} (IU) muons come from \( \nu _{\mu } \)
interactions in the lower apparatus \cite{mac31}. Since two scintillation counters
are intercepted, the T.o.F. is applied to identify the upward going
muons (Fig. \ref{fig2}). The average parent neutrino energy for these
events is 4.2 GeV (Fig. \ref{fig5}). If the atmospheric neutrino
anomalies were the results of \( \nu _{\mu }\rightarrow \nu _{\tau } \)
oscillations with maximal mixing and \( \Delta m^{2} \) between \( 10^{-3} \)
and \( 10^{-2}\, \, \eV ^{2} \), one would expect a reduction by about
a factor of two in the flux of these events, without any distortion
in the shape of the angular distribution. This is what is observed
in Fig. \ref{fig10}a.

The \textit{upstopping muons} (UGS) are due to external \( \nu _{\mu } \)
interactions yielding upgoing muons stopping in the detector.
The data correspond to an effective livetime of 5.6 y. The
\textit{semicontained downgoing muons} 
(ID) are due to \( \nu _{\mu } \)-induced
downgoing tracks with vertex in the lower MACRO (Fig. \ref{fig2}).
The two types of events are identified by means of topological criteria; the lack
of time information prevents to distinguish the two sub-samples. An
almost equal number of UGS and ID events is expected.  In case of oscillations with the quoted parameters,
the flux of the UGS should be reduced by 50\%, the
same amount of the ID muons at \( cos\theta \simeq -1 \).
No reduction is instead expected for the semicontained downgoing events
(coming from neutrinos having path lengths of \( \sim 20\km  \)).

MC simulations for the low energy data use the Bartol neutrino flux
and the neutrino low energy cross sections of ref. \cite{lipari94}.
The number of events and the angular distributions are compared with
MC predictions in Table \ref{tab:macro} and Figs. \ref{fig10}a,b. The
low energy data show a uniform deficit of the measured number of events
for the whole angular distribution with respect to predictions, \( \sim 50\% \)
for IU, 75\% for ID + UGS; there is good agreement with the predictions
based on neutrino oscillations with the parameters obtained from the
upthroughgoing muons.

\begin{table}
{\centering \begin{tabular}{cccc}
\hline 
                & Events    & MC-No oscillations  & $R = (Data/MC_{no osc}) $\\
                &           &                     &                        \\ \hline
Up throughgoing & $809$ & $1122 \pm 191$ & $(0.721 \pm 0.026_{stat}\pm 0.043_{sys}\pm 0.123_{th} )  $
\\ \hline
Internal Up     & $154$ & $285 \pm 28_{sys}\pm 71_{th}$ & $(0.54 \pm
0.04_{stat} \pm 0.05_{sys} 
\pm 0.13_{th} )$\\ \hline
Up Stop +  In Down &  262 & $ 375 \pm 37_{sys} \pm 94_{th} $ & $( 0.70 \pm
0.04_{stat}\pm 0.07_{sys} \pm 0.17_{th})$\\ \hline
\end{tabular}\par}

\caption{{\small Summary of the MACRO $\nu _{\mu}\rightarrow \mu$ events in
    $-1<cos \theta< 0$ after background subtraction.
For each topology (see Fig. \ref{fig2}) the number of measured events, the MC
prediction for no-oscillations and the ratio (Data/MC-no osc) are given.
}}

\label{tab:macro}
\end{table}

The average value of the double ratio \( R=(Data/MC)_{IU}/(Data/MC)_{ID+UGS} \)
over the measured zenith angle range is \( R\simeq 0.77 \pm 0.07 \); the
error includes statistical and theoretical uncertainties; \( R=1 \) is expected in case
of no oscillations \cite{mac54}.

\section{Search for Astrophysical Point Sources of High Energy Muon Neutrinos}

\begin{figure}
\begin{center}
\vspace{-3cm }
\hspace{-3cm }
  \mbox{ 
\epsfysize=6.6cm
         \epsffile{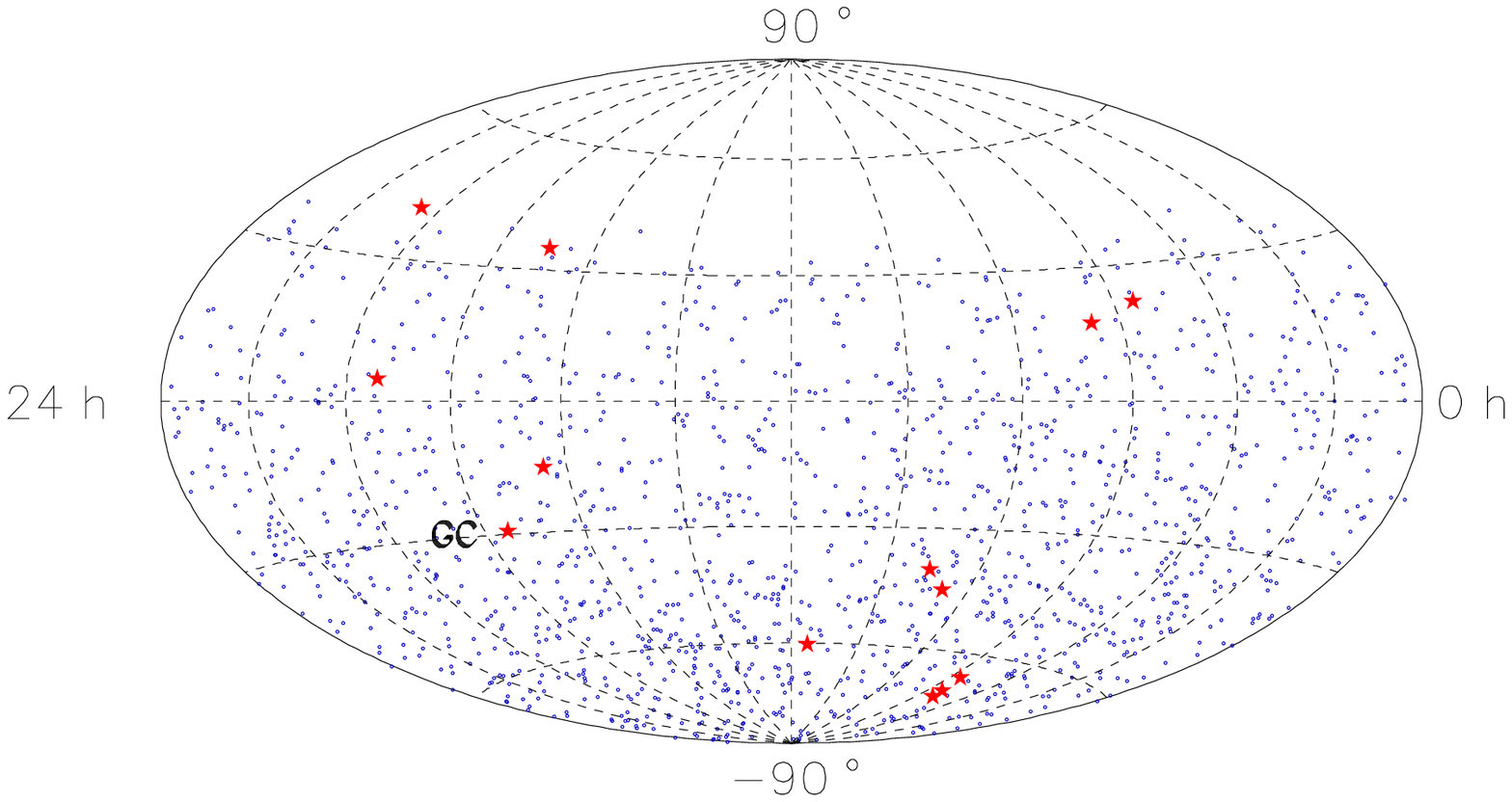}
\hspace{-2.1cm }
\epsfysize=8cm
         \epsffile{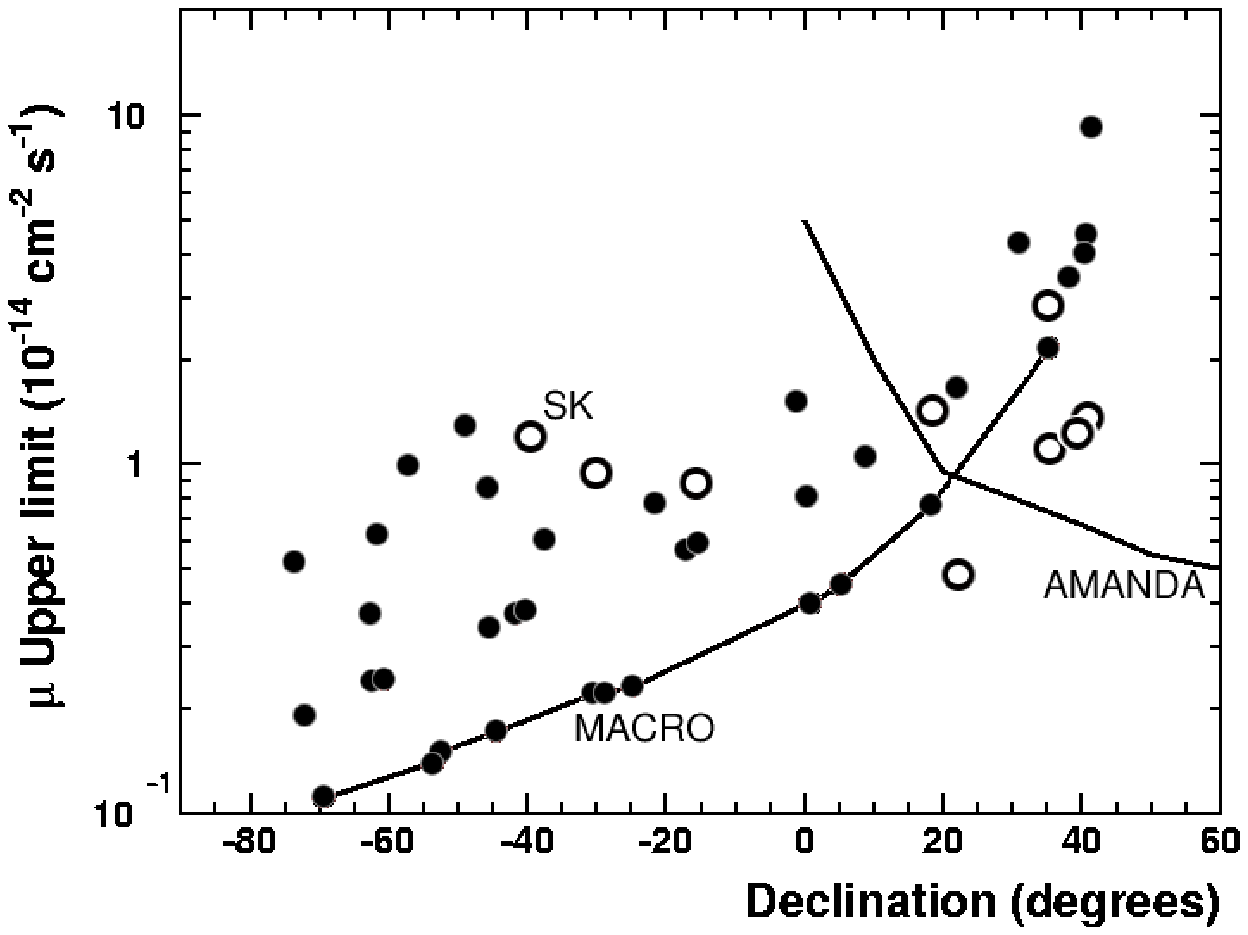} }
{\small \hskip 8.0 truecm (a) \hskip 7.5 truecm (b)}
\caption{\label{fig11}\small 
High energy neutrino astronomy. (a) Upgoing muon distribution in equatorial
coordinates (1356 events). 
(b) The black points are the MACRO 90 \% c.l. upwardgoing 
muon flux limits as a function 
of the declination for 42 selected sources. The solid line refers to the
MACRO limits obtained for those cases for which the atmospheric neutrino
background was zero. The limits from the SK (open circles) and AMANDA (thin
line) experiments are quoted; these last limits  refer to  much higher energy neutrinos.
}
\end{center}
\end{figure}

High energy \( \nu _{\mu } \) are expected to come from several
galactic and extragalactic sources. Neutrino production requires
astrophysical accelerators of charged particles and some kind of astrophysical beam
dumps. The excellent angular resolution of our detector allowed a
sensitive search for upgoing muons produced by neutrinos coming from
celestial sources, with a negligible atmospheric neutrino background.
An excess of events was searched for around the positions of known
sources in \( 3^{\circ } \) (half width) angular bins. This value
was chosen so as to take into account the angular smearing produced
by the multiple muon scattering in the rock below the detector and by the
energy-integrated angular distribution of the scattered muon, with
respect to the neutrino direction. Using a total livetime of 6.16
y (normalized to the complete configuration) we obtained a total of 1356 events,
see Fig. \ref{fig11}. The 90\% c.l. upper limits on the muon fluxes from
specific celestial sources lay in the range \( 10^{-15}-10^{-14}{\cm }^{-2}{\s }^{-1} \), see
Fig. \ref{fig11}b (preliminary data were reported at the 2001 conferences) \cite{mac37, mac39, mac48}.
The solid MACRO line is our sensitivity vs. declination. Notice that we
have two cases, GX339-4 ($\alpha = 255.71^o $, $\delta= -48.79^o $)  and Cir
X-1  ($\alpha = 230.17^o $, $\delta= -57.17^o $), with 7 events: in figure 11 they are
considered as a background, therefore the upper flux limits
are higher; but they could also be indication of signals.

We searched for time coincidences of our upgoing muons with \( \gamma  \)-ray
bursts as given in the BATSE 3B and 4B catalogues, for the period
from April 91 to December 2000 \cite{mac33}. No statistically significant
time correlation was found.

We have also searched for a diffuse astrophysical neutrino flux for
which we establish a flux upper limit at the level of \( 1.5\cdot
10^{-14}{\cm }^{-2}{\s }^{-1} \) \cite{mac48}.

\section{Indirect Searches for WIMPs}

Weakly Interacting Massive Particles (WIMPs) could be part of the
galactic dark matter; WIMPs could be intercepted by celestial bodies,
slowed down and trapped in their centers, where WIMPs and anti-WIMPs could
annihilate and yield upthroughgoing muons. The annihilations in these
celestial bodies would yield  neutrinos of \GeV{} or \TeV{} energy,
in small angular windows from their centers. 

\begin{figure}
\vspace{-1.cm}
 \begin{center}
  \mbox{ \epsfysize=6cm
         \epsffile{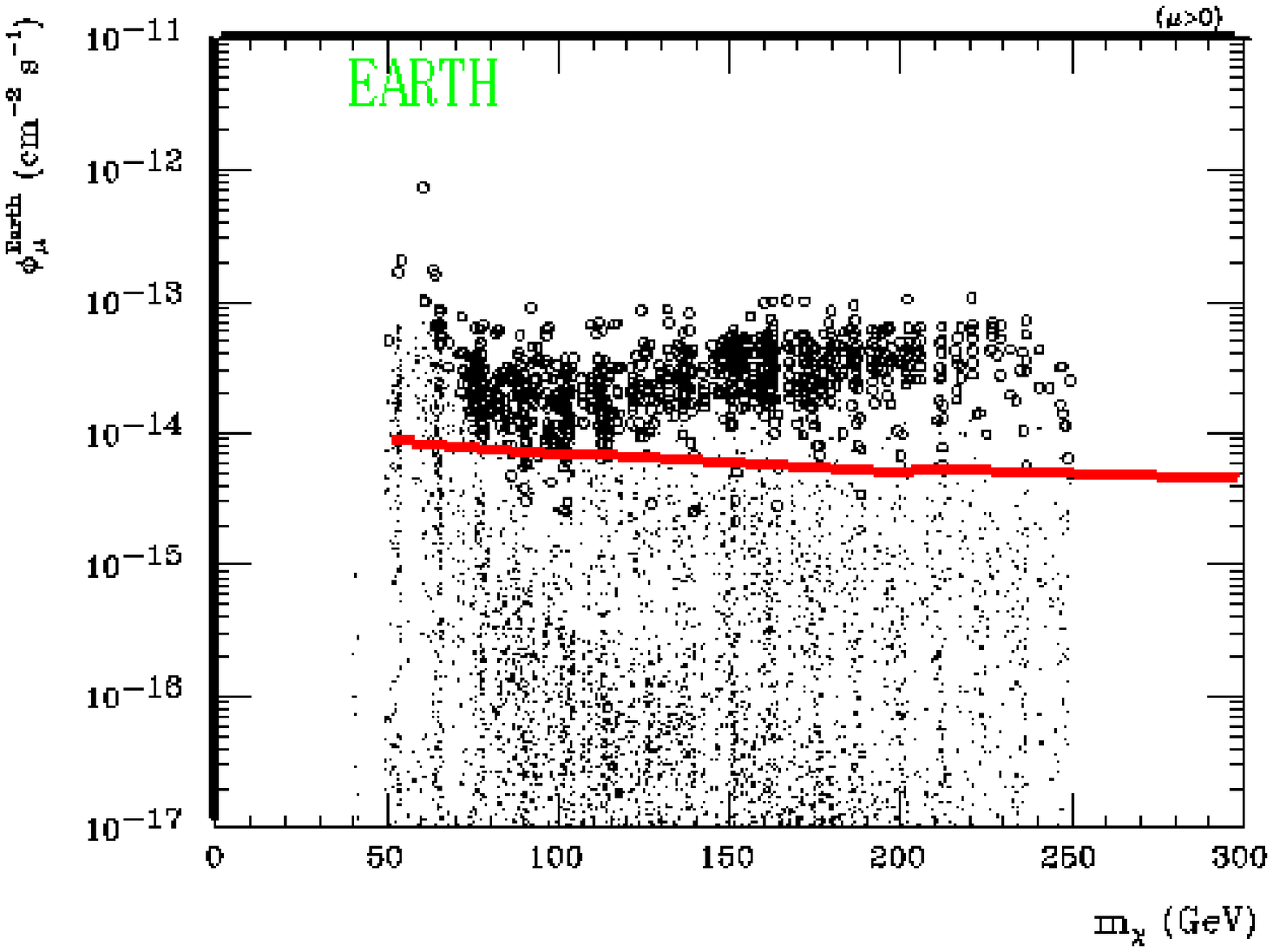}
\epsfysize=6cm
         \epsffile{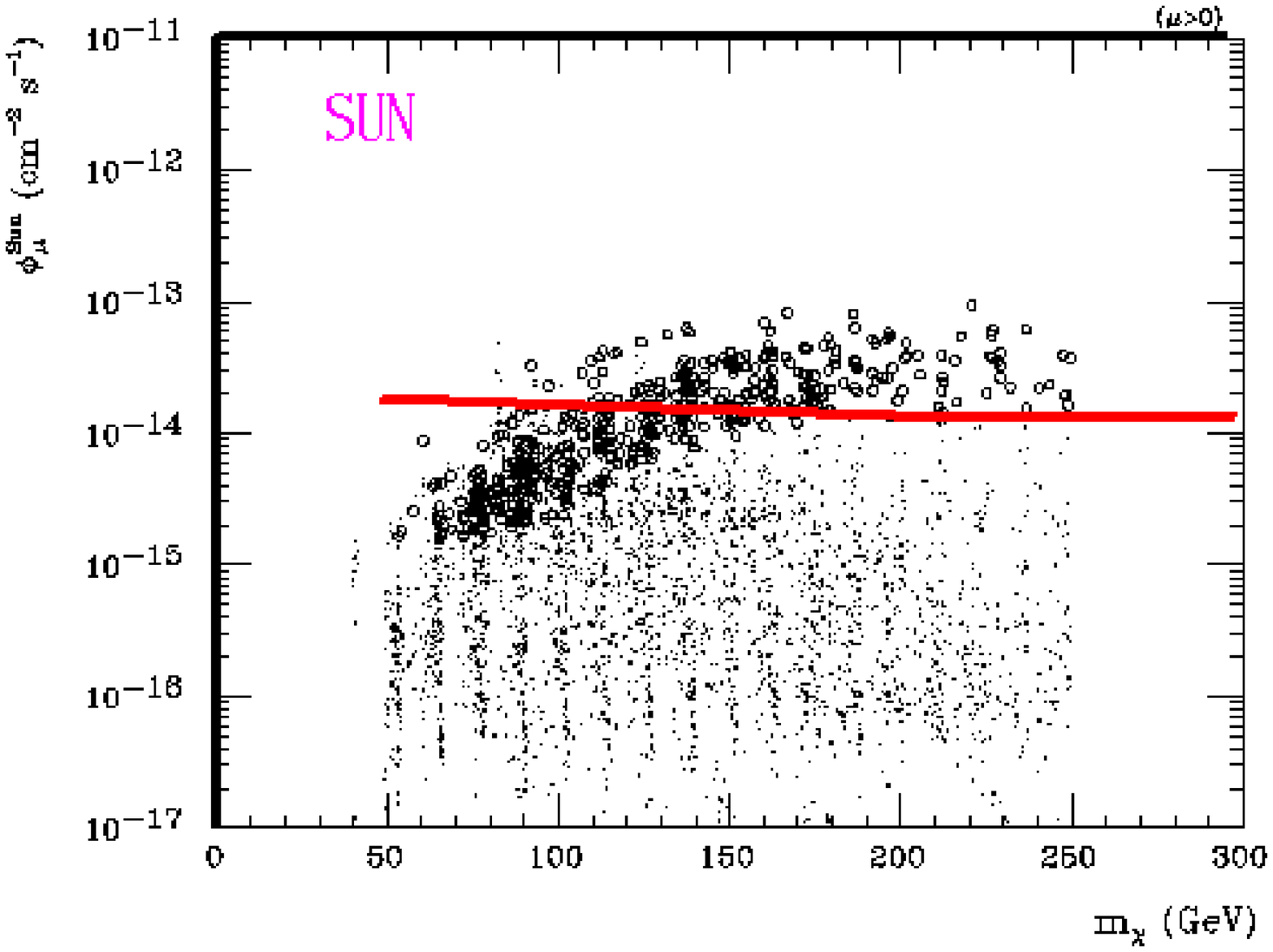}   }
 \end{center}
{\centering 
{\small \hskip 5.0 truecm (a) \hskip 7.5 truecm (b)}}
\caption{\label{fig12}\small  
(a) The solid line is the MACRO upwardgoing muon flux upper limit (90\%c.l.) from the Earth plotted
vs. neutralino mass $m_\chi$ for ${E_\mu }^{th}=1 \GeV$.  (b) The same as
in (a) but for upwardgoing muons from the Sun \cite{mac28}.
Each dot is obtained varying model parameters. The open circles indicate
models {\it excluded  } by direct measurements (in particular the DAMA/NaI experiment 
\cite{dama}) and assume a local dark matter density of about $ 0.5 \GeV cm^{-3}$.}
\end{figure}

For the Earth we have chosen a \( 15^{o} \) cone around the vertical:
we find 863 events. The MC expectation for atmospheric \( \nu _{\mu } \)
without oscillations gives a larger number of events. We set a conservative
flux upper limit assuming that the measured number of events equals the
expected ones. We obtain the \( 90\, \% \) c.l. MACRO limits for
the flux of upgoing muons as shown in Fig.  \ref{fig12}a (it varies from about 0.8
to 0.5 \( 10^{-14}{\cm }^{-2}{\s }^{-1} \)). If the WIMPs are identified
with the smallest mass neutralino, the MACRO limit may be used to
constrain the stable neutralino mass, following the model of Bottino
et al. \cite{bottino}, see Figure \ref{fig12}a.

A similar procedure was used to search for muon neutrinos from the
Sun, using 10 search cones from \( 3^{o} \) to \( 30^{o} \). In
the absence of statistically significant excesses the muon upper limits
are at the level of about \( 1.5-2\cdot 10^{-14}{\cm }^{-2}{\s }^{-1} \) are
established. The limits are shown in Fig.  \ref{fig12}b as a function of the WIMP (neutralino) mass.

\section{Magnetic Monopoles and Nuclearites}

The search for magnetic monopoles (MM) was one of the main objectives
of our experiment. Supermassive monopoles predicted by Grand Unified
Theories (GUT) of the electroweak and strong interactions should have 
masses of the order of \( m_M\sim 10^{17}{\GeV } \). 

These monopoles could be present in the penetrating cosmic radiation and
are expected to have typical galactic velocities, \( \sim 10^{-3}c \),
if trapped in our Galaxy. MMs trapped in our solar system or in the supercluster of galaxies
may travel with typical velocities of the order of \( \sim 10^{-4}c \)
and \( \sim 10^{-2}c \), respectively. Monopoles in the presence
of strong magnetic fields may reach higher velocities. Possible  intermediate
mass monopoles could reach relativistic velocities.

The reference sensitivity level for a significant MM search is the
Parker bound \cite{giapat}, the maximum monopole flux compatible with the survival
of the galactic magnetic field. This limit is of the order of \( \Phi \lsim 10^{-15}{\cm }^{-2}{\s }^{-1}{\sr }^{-1} \),
but it could be reduced by almost an order of magnitude when considering
the survival of a small galactic magnetic field seed \cite{giapat}.
Our experiment was designed to reach a flux sensitivity well below
the Parker bound, in the MM velocity range of \( 4\times 10^{-5}<\beta <1. \)
The three MACRO sub-detectors have sensitivities in wide \( \beta  \)-ranges,
with overlapping regions; thus they allow multiple signatures of the
same rare event candidate. No candidates were found in several years
of data taking by any of the three subdetectors.

\begin{figure}
\vspace{-1.cm}
 \begin{center}
  \mbox{ \epsfysize=10cm \epsffile{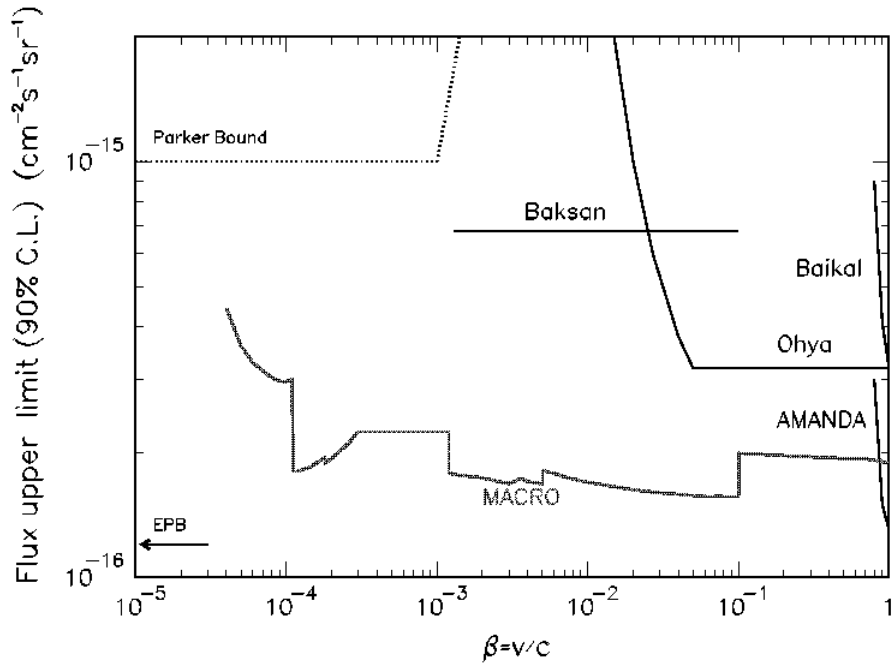}}
 \end{center}
\vspace{-.5cm}
\caption{\label{fig13}\small  
Magnetic monopole flux upper limits at the $90\%$ c.l. obtained by MACRO and by other experiments 
\cite{giapat,nucl1,nucl2}. The limits apply to singly charged ($ { g =g_D }$) monopoles
assuming that catalysis cross sections are smaller than a few mb.}
\end{figure}

The MM flux limits set by several different analyses using the three subdetectors
over different \( \beta  \)-range were combined to obtain a global
MACRO limit. For each \( \beta  \) value, the global time integrated
acceptance was computed as the sum of the independent portions of
each analysis. Our limits are shown in Fig. \ref{fig13} versus \( \beta  \)
together with the limits set by other experiments \cite{giapat, nucl1, nucl2}; other limits 
are quoted in \cite{mac36, mac38, mac40, mac49, bakari}. 
Our MM limits are the best existing over a wide range of \( \beta  \) , \( 4\times 10^{-5}<\beta <1. \)

A specific search for monopole catalysis of nucleon decay was made with the
streamer tube system \cite{mac47}. Since no event was found, we can place a
monopole flux upper limit at the level of $\sim 3 \cdot 10^{-16}
\,cm^{-2}s^{-1}sr^{-1}$ for $10^{-4} \lsim \beta \lsim 5 \cdot 10^{-3}$,
valid for a large catalysis cross section, $5\cdot 10^2 < \sigma_{cat}
< 10^3\, mb$. The flux limit for the  standard direct MM search with
streamer tubes is valid for  $\sigma_{cat} < 100\,mb$.

\begin{figure}
\vspace{-1.cm}
 \begin{center}
  \mbox{ \epsfysize=8.5cm
         \epsffile{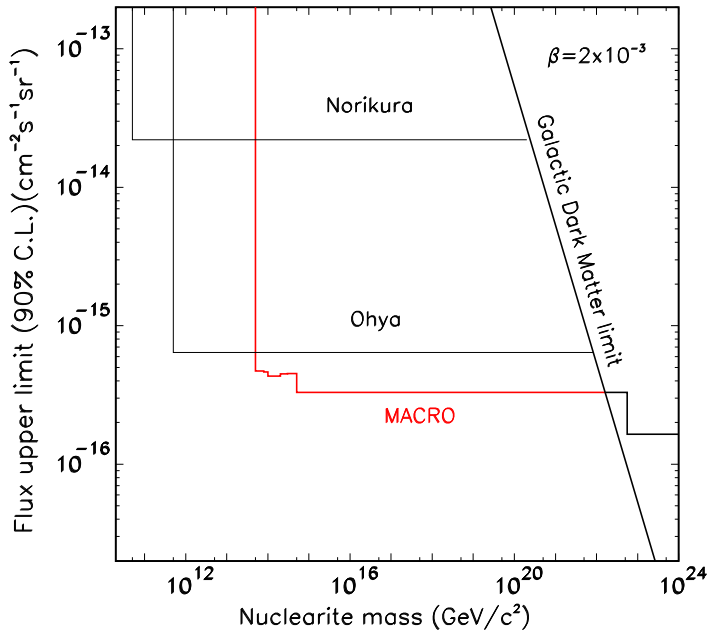}   }
 \end{center}
\caption{\label{fig14}\small  
$90 \%$ c.l. flux upper limits vs. mass for nuclearites with $\beta = 2 \cdot 10 ^ {-3}$ at ground level. Nuclearites of such velocity could have galactic or
extragalactic origin. The MACRO direct limit (solid line) is shown along
with the other direct limits 
\cite{nucl1,nucl2}; the indirect mica limits of
\cite{nucl3,nucl4} are at the level of $2 \cdot 10^{-20 } cm^{-2}  s^{-1}  sr^{-1} $. The limits for nuclearite masses larger than $ 5 \cdot 10^ {22 }\GeV c^ {-2 }$ correspond to an isotropic flux.
}
\end{figure}

The MM searches based on the scintillator and on the nuclear track subdetectors
were also used to set new upper limits on the flux of cosmic ray nuclearites
(strange quark matter), over the same \( \beta  \) range, Fig. \ref{fig14}.
If nuclearites are part of the dark matter in our galaxy, the most
interesting \( \beta  \) is of the order of \( \sim 10^{-3} \).
Fig. \ref{fig14}  shows our limit plotted vs nuclearite mass for \( \beta =2\times 10^{-3} \) (at 
ground level).
Other experimental limits are also shown. 

Some of the nuclearite limits apply also to Q-balls (agglomerates
of squarks, sleptons and Higgs fields) \cite{mac47, mac49}. 

The energy losses of MMs, dyons and of other heavy particles in the 
Earth and in different detectors for various particle masses and 
velocities were computed in \cite{derka}.

\section{Neutrinos from Stellar Gravitational Collapses}

A stellar gravitational collapse (GC) of the core of a massive star
is expected to produce a large burst of all types of neutrinos and
antineutrinos with energies of \( 7-30 \)~\MeV{} and with a duration
of \( \sim 10{\s } \). The \anue{}'s can be detected via the process
\( \bar{\nu }_{e}+p\rightarrow n+e^{+} \) in the liquid scintillator.
About \( 100\div 150 \) \anue{} events should be detected in our
580 t scintillator for a stellar collapse at the center of our Galaxy.

We used two electronic systems for detecting \anue{}'s from stellar
gravitational collapses. The first system was based on the dedicated
PHRASE trigger, the second one was based on the ERP trigger. Both
systems had an energy threshold of \( \sim 7{\MeV } \) and recorded
pulse shape, charge and timing informations. Immediately after a \( >7{\MeV } \)
trigger, the PHRASE system lowered its threshold to about 1 MeV, for
a duration of \( 800{\mu \s } \), in order to detect (with a \( \simeq 25\, \% \)
efficiency) the \( 2.2{\MeV } \) \( \gamma  \) released in the reaction
\( n\, +\, p\rightarrow d\, +\, \gamma _{2.2\MeV } \) induced by
the neutron produced in the primary process.

A redundant supernova alarm system was in operation, alerting immediately
the physicists on shift. We defined a general procedure to alert the
physics and astrophysics communities in case of an interesting alarm. Finally, a procedure to link the various supernova
observatories around the world was set up \cite{mac23}.

The effective MACRO active mass was \( \sim 580\, \,  \)t; the live-time fraction
in the last four years was \( \simeq 97.5\, \% \). No stellar gravitational
collapses were observed in our Galaxy from the beginning of \( 1989 \)
to the end of 2000 \cite{mac53}.

\section{Cosmic Ray Muons}

The large area and acceptance of our detector allowed the study of
many aspects of physics and astrophysics of cosmic rays (CR). We recorded
 \( \sim 6\times 10^{7} \) single muons and \( \sim3.7 \times 10^{6} \)
multiple muons at the rate of \( \sim 18,000 \)/day.

\noindent \textbf{Muon vertical intensity.} The underground muon vertical
intensity vs. rock thickness provides information on the high energy
(\( E\, \, \gsim \, \, 1.3{\TeV } \)) atmospheric muon flux and on
the all-particle primary CR spectrum. The results can be used to constrain
the models of cosmic ray production and interaction. The analysis
performed in \( 1995 \) covered the overburden range \( 2200\div 7000\hg /\cm ^{2} \)
\cite{mac16}.

\begin{figure}
\vspace{-1cm}
  \begin{center}
  \mbox{ 
\hspace{-1cm}
 \epsfysize=9cm
         \epsffile{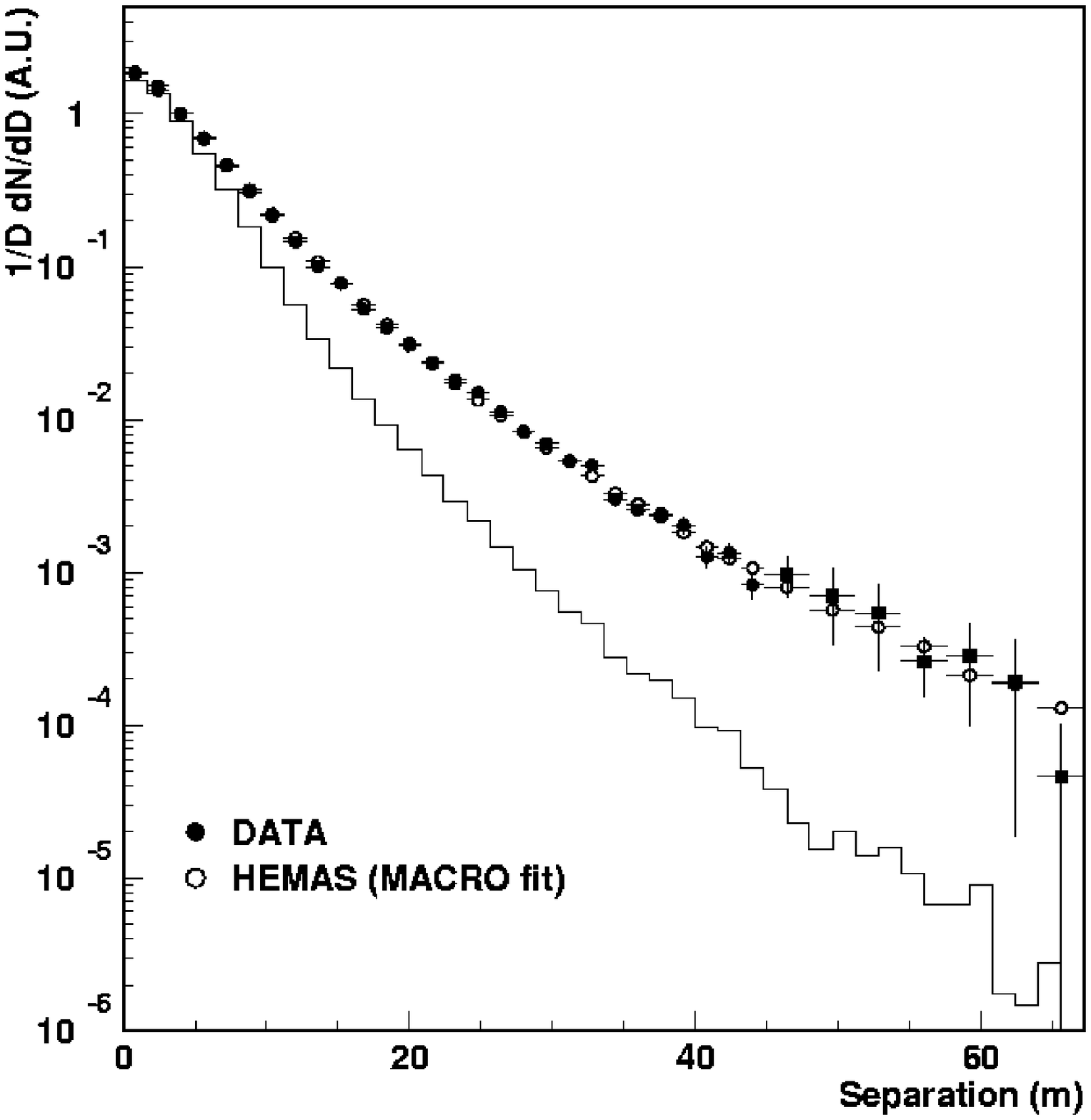} 
 \epsfysize=9cm
         \epsffile{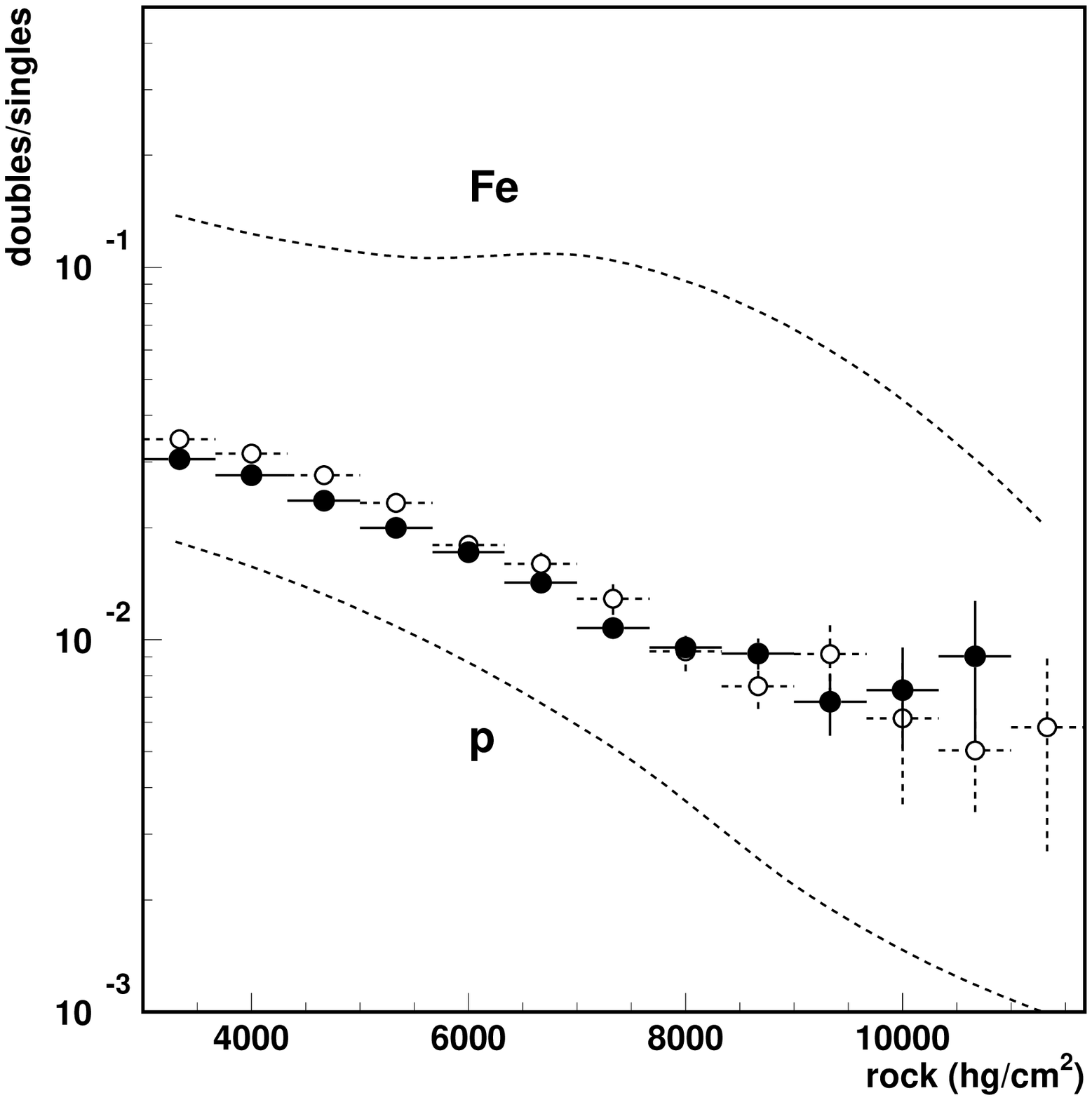} 
}
{\small \hskip 4.0 truecm (a) \hskip 7.5 truecm (b)}   
\caption{\label{fig15}\small (a) True unfolded experimental decoherence 
distribution for an infinite detector (black points) compared with MC 
expectations (open points); the measured decoherence distribution before 
unfolding is shown as an histogram \cite{mac29}.
(b) Ratio of double muon events to single muon events as a function
of the rock depth. The black points are our data; the open circles are
Monte Carlo predictions made using the MACRO composition model. Monte Carlo predictions using
pure proton and iron primaries are shown as dashed lines.}
\end{center}
\end{figure}

\noindent \textbf{Analysis of high multiplicity muon bundles.} The
study of the \textbf{multiplicity distribution} of muon bundles provides
informations on the primary CR composition model. 
The study of the \textbf{decoherence} function (the distribution of the distance
between two muons in a muon bundle) provides informations on the hadronic interaction
features at high energies; this study was performed using a large sample
of data and improved Monte Carlo methods, see Fig. \ref{fig15}a \cite{mac29}.
We used different hadronic interaction 
models (DPMJET, QGSJET, SIBYLL,
HEMAS, HDPM) interfaced to the HEMAS and CORSIKA shower propagation
codes \cite{tesiphd}.

We studied \emph{muon correlations inside a bundle} \cite{tesiphd}, 
using the so called correlation integral 
\cite{corint}, to search for correlations of dynamical origin in the
bundles.  Since the cascade development in atmosphere is mainly determined by the number of {}``steps{}''
in the {}``tree formation{}'', we expect a different behaviour for cascades originated by  light and
heavy CR primaries. For the same reason, the analysis
should be less sensitive to the hadronic interaction model adopted
in the simulations. This analysis shows that, in the
energy region above 1000 TeV, the composition model derived from the
analysis of the muon multiplicity distribution \cite{mac19, mac20}
is almost independent from the interaction model.

\par We also  searched for  \emph{substructures ({}``clusters{}'')
inside muon bundles} \cite{clusters}. The search for clusters
was performed by means of different software algorithms; the study
is sensitive both to the hadronic interaction model and to the primary CR
composition model. 
If the primary composition has been determined by the first method, a choice of the bundle topology
gives interesting connections with the early hadronic interaction
features in the atmosphere.  The comparison between our data and Monte
Carlo simulations allowed to place constraints on  the used interaction models. The same Monte Carlo study has
shown that muon bundles with a central core and an isolated cluster
with at least two muons are the result of random associations of peripheral
muons. A combined analysis  with the study of the decoherence
function for high multiplicity events has shown that the hadronic
interaction model that better reproduces the underground observables
is  QGSJET.\\
\textbf{The ratio double muons/single muons:} The ratio $N_2$/$N_1$ of double muon events over
single muon events is expected to decrease at increasing  rock
depths, unless some exotic phenomena occur. The ratio $N_2$/$N_1$
was studied in  underground experiments and in phenomenological
papers \cite{elbert, ryazhaskaya}. The LVD collaboration reports that
the ratio of multiple-muons to all-muons increases  for rock depths $h> 7000~hg/cm^2$.

We measured \cite{mac44}  the ratio $N_2$/$N_1$ as a function of the rock depth, using also multiple muon events at
large zenith angles. A detailed Monte Carlo simulation was made using the HEMAS code, where the event zenith
angle can be extended up to $89^\circ$.
The event direction
is reconstructed by the tracking system. The
rock depth is provided by the Gran Sasso map function $h(\theta,\phi)$,
which extends up to $\theta =94^\circ$.  The ``true'' muon multiplicity is the largest value among $N_{HW}$ and
$N_{VW}$, the multiplicities in the horizontal and vertical planes,
respectively. Monte Carlo simulations have shown that the percentage
of events with mis-reconstructed multiplicity is less that 3\%. Attention was 
devoted to the ``cleaning'' of the events from spurious effects (electronic
noise, radioactive background, etc); in many cases, we made a
visual scanning of the events.
Our measured ratios $N_{2}$/$N_{1}$ as a function of the rock depth,
Fig. \ref{fig15}, are 
in agreement with the expectation of a monotonic decrease
of $N_{2}$/$N_{1}$ down to $h~\sim~10000~hg/cm^{2}$. Above this value,
the low statistics does not allow to state a firm conclusion
on a possible increase of $N_{2}$/$N_{1}$.\\
\textbf{Muon Astronomy.} In the past, some experiments reported 
excesses of modulated muons from the direction of known astrophysical
sources, f.e. Cyg X-\( 3 \). Our data do not indicate
significant excesses above  background, both for steady
dc fluxes and for modulated ac fluxes.
\begin{figure}
  \begin{center}
  \mbox{ 
\epsfysize=8cm
         \epsffile{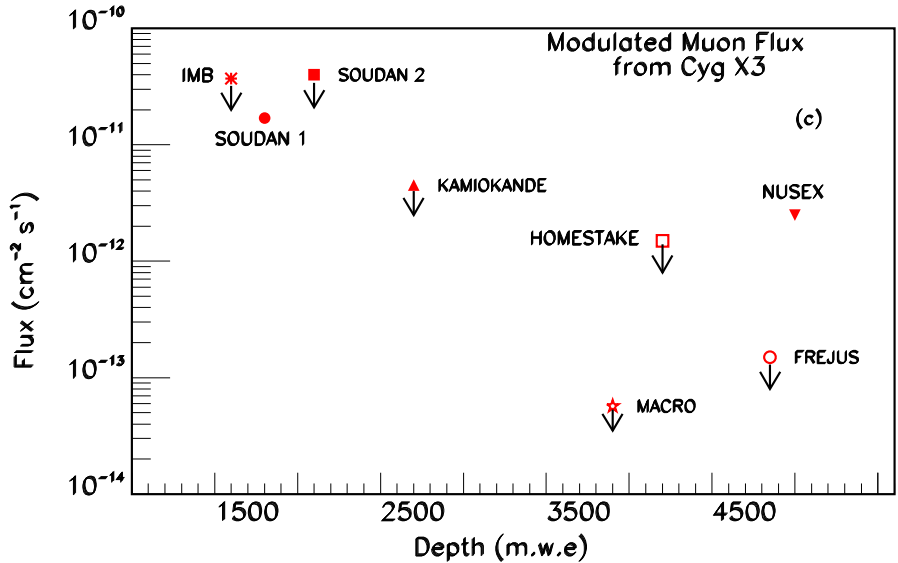} 
}   
\caption{\label{fig16}\small Present situation of the searches for a
  modulated muon signal from Cyg X-3. The Soudan 1 and Nusex collaborations
  reported positive indications, while all other experiments give flux
  upper limits.}

\end{center}
\end{figure}
The MACRO pointing precision was checked via the shadow of the Moon
and of the Sun on primary cosmic rays.  The pointing resolution
was checked with double muons, assuming that they are parallel. The  angle
containing  \( 68\% \) of the events in a \( \Delta \theta  \) bin
is \( 0.8^{o} \), which we take as our resolution.

\begin{figure}
  \begin{center}
  \mbox{ 
\epsfysize=10cm
         \epsffile{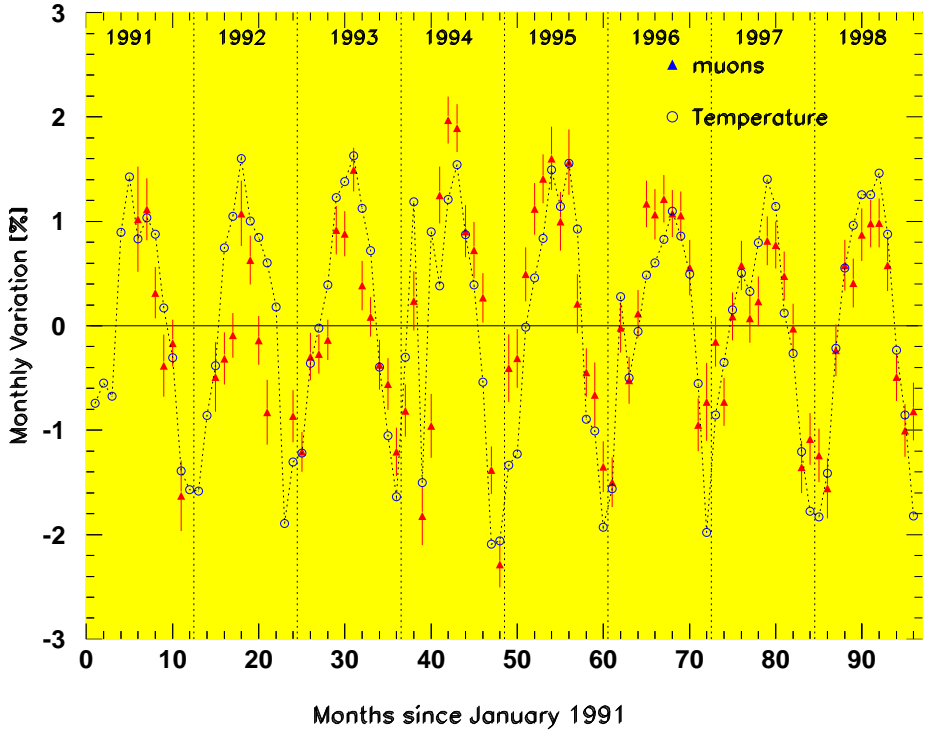} 
}   
\caption{\label{fig17}\small 
Seasonal variation of the muon flux from above (black triangles); the open circles are measurements of the temperature of the upper atmosphere.}
\end{center}
\end{figure}

\begin{figure}
\vspace{-1.cm}
 \begin{center}
  \mbox{ \epsfysize=8cm
         \epsffile{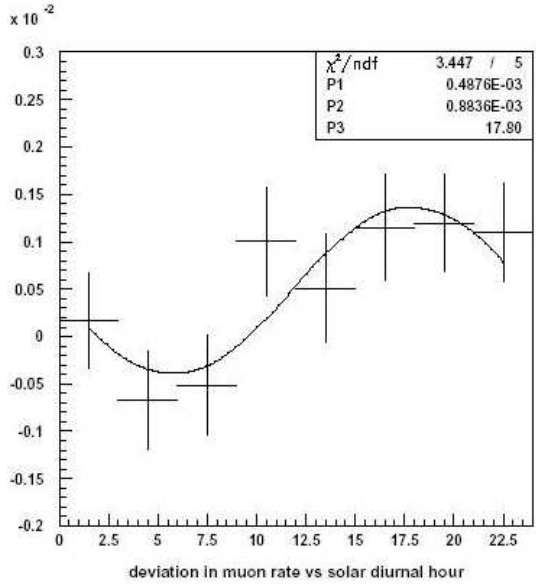}   
\epsfysize=8cm
         \epsffile{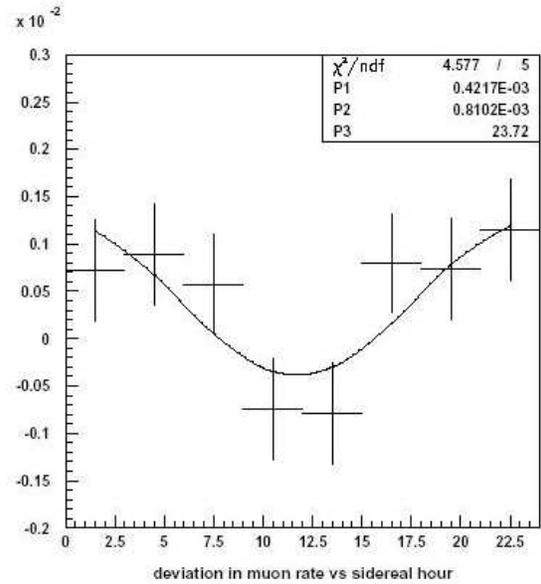}   }
 \end{center}
{\centering 
{\small \hskip 4.0 truecm (a) \hskip 7.5 truecm (b)}}
\caption{\label{fig18}\small  
Deviations of the muon rate from the mean muon rate (a) versus the local
solar diurnal time at  Gran Sasso, and (b) versus the local sidereal time.
}
\end{figure}

\noindent \textit{All sky d.c. survey.} The sky, in galactic coordinates,
was divided into bins of equal solid angle, \( \Delta \Omega =2.1 \cdot 10^{-3}sr \),
\( \Delta \alpha =3^{o} \), \( \Delta sin\delta =0.04 \); they correspond
to  cones of \( 1.5^{o} \) half angles. In order to remove
edge effects, three other surveys were done, by shifting the map by
one-half-bin in \( \alpha  \) (map 2), by one-half bin in sin\( \delta  \)
(map 3) and with both \( \alpha  \) and sin\( \delta  \) shifted
(map 4). For each solid angle bin we computed the deviation from the
average measured muon intensity, after background subtraction, in units
of standard deviations 
\begin{equation}
\sigma (i)={{N_{obs}(i)-N_{exp}(i)}\over {\sqrt{N_{exp}(i)}}}
\end{equation}
 where \( N_{obs}(i) \) is the observed number of events in bin \textit{i}
and \( N_{exp} \) is the number of events expected in that bin from
the simulation. No deviation was found and for the majority
of the bins we obtain flux upper limits at the level of  \(
{\Phi _{\mu }}^{steady}(95\%)\leq 5\times 10^{-13}cm^{-2}s^{-1}\).\\
\textit{Specific point-like d.c. sources}. For 
Cyg X-3, Mrk421, Mrk501 we searched in a narrower cone (\( 1^{o} \)
half angle) around the source direction. We obtain flux limits at
the level of \( (2-4)\cdot 10^{-13}cm^{-2}s^{-1} \). There is a
small excess of \( 2.0\, \, \sigma  \) in the direction
of Mrk501.\\
 \textit{Modulated a.c. search from Cyg X-3 and Her X-1}. No evidence
for an excess was observed and the limits are \( {\Phi <2\times 10^{-13}cm^{-2}s^{-1}} \);
see Fig. \ref{fig16} .\\
 \textit{Search for bursting episodes}.  We made a search for pulsed muon signals in a 
\( 1^{o} \) half angle cone around the location of  possible sources
of high energy photons. Bursting episodes of duration of
\( \sim 1 \) day were searched for with two different methods. In the
first method we searched for daily excesses of muons above the background,
also plotting cumulative excesses day by day. In the second method
we computed day by day the quantity \( -Log_{10}P \) where \textit{P}
is the probability to observe a burst at least as large as \( N_{obs} \).
We find some possible excesses for Mrk421 on the days 7/1/93, 14/2/95,
27/8/97, 5/12/98.  \\
 \textit{Seasonal variations.} Underground muons are produced by mesons
decaying in flight in the  atmosphere. The muon flux thus depends
on the ratio between the decay and the interaction probability of
the parent mesons, which are sensitive to the atmospheric density
and to the average temperature. The flux is expected to decrease in
winter, when the temperature is lower and the atmosphere more dense,
and to increase in summer. We find the expected variations at the
level of \( \pm 2\% \) \cite{mac21}, see Fig. \ref{fig17}.\\
\textit{Solar daily variations.} Because of variations in the day-night
temperatures we expect solar daily variations similar to seasonal
variations, but of considerably smaller amplitudes. Using the total
MACRO data,  we find these variations with an amplitude $A = (0.88 \pm 0.26) \cdot 10^{-3}$ with
a significance of about 3.4 $\sigma$, see Fig.  \ref{fig18}a.

\begin{figure}
\vspace{-1.cm}
 \begin{center}
  \mbox{ \epsfysize=8cm
         \epsffile{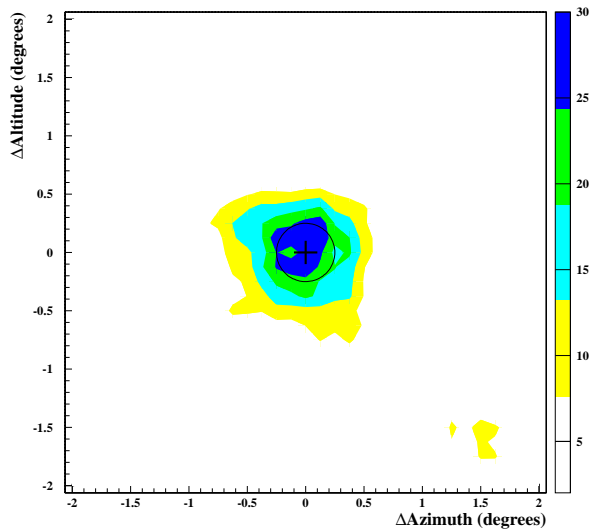}
\epsfysize=8cm
         \epsffile{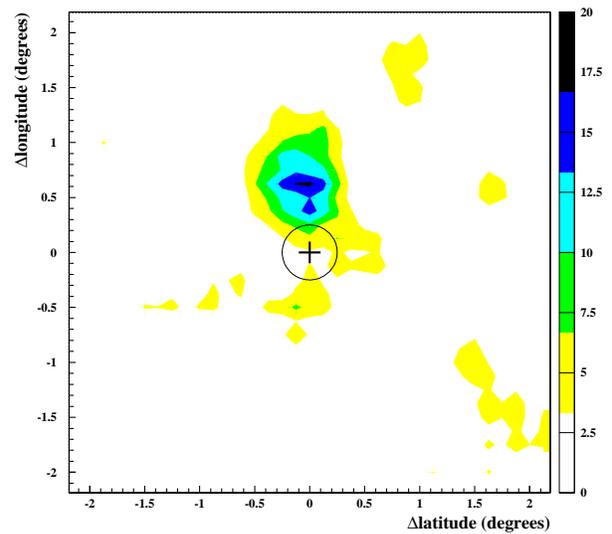}   }
 \end{center}
{\small \hskip 4.0 truecm (a) \hskip 7.5 truecm (b)}
\caption{\label{fig19}\small  
Moon and sun shadows. (a) Two dimensional distributions of the muon event
density around the moon direction. The various regions of increasing gray
scale indicate various levels of deficit in percent. The darkest one
corresponds to the maximum deficit.  
(b) Same analysis for the sun direction.
}
\end{figure}
\begin{figure}
  \begin{center}
\vspace{-1.cm}
  \mbox{ 
\epsfysize=9.5cm
         \epsffile{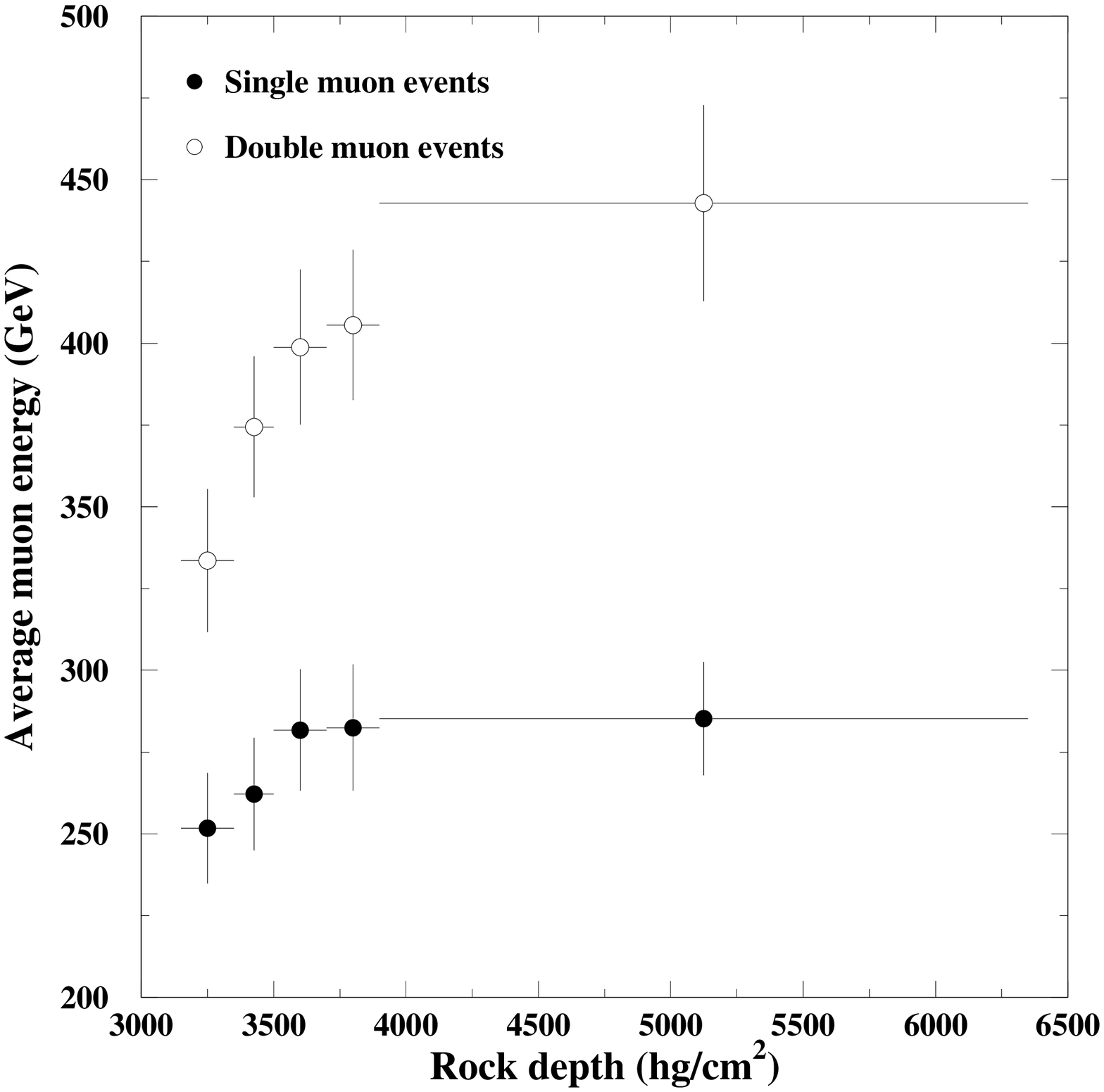} 
}   
\caption{\label{fig20}\small Residual average muon energy at the underground Gran
  Sasso lab versus standard rock depth for single muons and for double
  muons \cite{mac27}, see text.
}
\end{center}
\end{figure}

\noindent \textit{Sidereal anisotropies}  are
due to the motion of the solar system through the  ``sea''  of relativistic cosmic rays
in our galaxy. They are expected to yield a small effect.  
After a correction due to the motion of the Earth around the Sun, we
observe 
variations with  an amplitude of 
$8.6 \cdot 10^{-4}$ and a phase $\phi_{max} = 22.7^\circ$
with a statistical significance of 3 $\sigma$, Fig.  \ref{fig18}b.\\
\textbf{Moon and sun Shadows of primary cosmic rays.} The pointing
capability of MACRO was demonstrated by the observed ``shadows''
of the Moon and of the Sun, which produce a \lq\lq shield\rq\rq~to
the cosmic rays. We used a sample of \( 45\cdot 10^{6} \) muons,
looking at the bidimensional density of the events around the directions
of the Moon and of the Sun \cite{mac26}\cite{mac45}. In Fig. \ref{fig19}
we show two-dimensional plots of the muon deficits caused by the Moon
and the Sun. For the Moon: we looked for events in a window \( 4.375^{o}\times 4.375^{o} \)
centered on the Moon; the window was divided into \( 35\times 35 \)
cells, each having dimensions of \( 0.125^{o}\times 0.125^{o} \)
(\( \Delta \Omega =1.6\cdot 10^{-2}deg^{2} \)). In the bidimensional
plot of Fig. \ref{fig19}a one observes a depletion of events with
a statistical significance of \( 5.5\, \, \sigma  \). The observed
slight displacement of the maximum deficit is consistent with the
displacement of the primary protons due to the geomagnetic field.
We  repeated the same analysis for muons in the sun window, Fig.
\ref{fig19}b. The  difference between the apparent sun position
and the observed muon density is due to the combined effect of the
magnetic field of the Sun and of the Earth. The observed
depletion has a statistical significance of \( 4.5\, \, \sigma  \).
\section{Muon energy measurement with the TRD detector }
The underground differential energy spectrum of muons was measured
with the three TRD modules detector. We analyzed two types of
events: ``single muons'', i.e. single events in MACRO
crossing a TRD module, and ``double muons'', i.e.
double events in MACRO with only one muon crossing the TRD detector.The
measurements refer to muons with energies \( 0.1<E_{\mu }<1\TeV  \)
and for \( E_{\mu }>1\TeV  \) \cite{mac27}. In order to evaluate
the local muon energy spectrum, we must take into account the TRD
response function, which induces some distortion of the ``true''
muon spectrum distribution. The ``true'' distribution
was extracted from the measured one by an unfolding procedure that
yields good results only if the response of the detector is correctly
understood. We used an unfolding technique developed according
to Bayes' theorem. Fig. \ref{fig20} shows the average muon energies
versus standard rock thickness for single and double muons. Systematic
uncertainties are included in the error bars. The average single muon
energy at the Gran Sasso underground lab is \( 270\, \, \GeV  \);
for double muons it is \( \sim 380\, \, \GeV  \). Double muons are more
energetic than single muons; this is in agreement with the predictions
of interaction models of primary CRs in the atmosphere.

\section{EAS-TOP/MACRO Coincidence Experiment }
For coincident events, EASTOP  measured the e.m. size of the showers above
the surface (at Campo Imperatore), while MACRO measured penetrating muons underground.
The purpose is to study the primary cosmic ray composition versus energy
reducing the dependence on the interaction and propagation models.
The two completed detectors  operated in coincidence for a livetime of  960.1 days.
The number of coincident events is 28160, of which 3752
have shower cores inside the edges of the EASTOP array (``internal events'')
and shower sizes $N_e > 2 \cdot 10^5$;  409 events have $N_e > 10^{5.92}$, i.e.
above the CR knee.
The data  have been analyzed in terms of the QGSJET 
interaction model as implemented in CORSIKA \cite{kna5}. 

The e.m. detector of EASTOP is made of 35 scintillator modules, 10 m$^2$
each, covering an area of  $\simeq 10^5 \m^2$.
The array is fully efficient for $N_e > 10^5$.
The  reconstruction capabilities of the extensive air shower (EAS) parameters 
for internal events are:
${{\Delta N_e} \over N_e} \simeq 10 \%$ for $N_e \aprge 10^5$,
and $\Delta \theta \sim 0.9^o$ for the EAS arrival direction \cite{as2}.

We considered in MACRO muon tracks with at least 4 aligned hits in both views of the horizontal streamer tube planes.
The muon energy threshold at the surface inside the effective area of EAS-TOP, for muons reaching the MACRO depth,
ranges from  1.3 TeV to  1.8 TeV.   
Event coincidence is made off-line, using the absolute time given by
a GPS system with an accuracy of better than 1 $\mu s$.
Independent analyses of the MACRO and of the EAS-TOP data are described in  \cite{mac20} and \cite{as5}, respectively.

The main experimental features considered are the muon multiplicity
distributions in six different intervals of shower sizes. For each size
bin the muon multiplicity 
distribution was fitted with a superposition of (i) pure $p$ and $Fe$
components, or (ii) light $(L)$ and
heavy $(H)$ admixtures containing equal fractions of $p$ and $He$ or $Mg$ and 
$Fe$, respectively.  
All spectra in the simulation have slope $\gamma = 2.62$. In each of the
six windows we minimized

\begin{equation}
\chi^2 = \sum_i{\frac{(N^{exp}_i - p_1 N^{p}_i - p_2 N^{Fe}_i)^2
}{\sigma_{i,exp}^2}} 
\end{equation}
where $N^{exp}_i$ is the number of observed events in the $i$-th bin,
$N^p$ ($N^L$) and $N^{Fe}$ ($N^H$) 
are the number of simulated events in the same  bin, 
$p_1$ and $p_2$ are the parameters (to be fitted) defining the
fraction of each mass component in  the same multiplicity bin.

\begin{figure}
  \begin{center}

  \mbox{ 
\epsfysize=8cm
         \epsffile{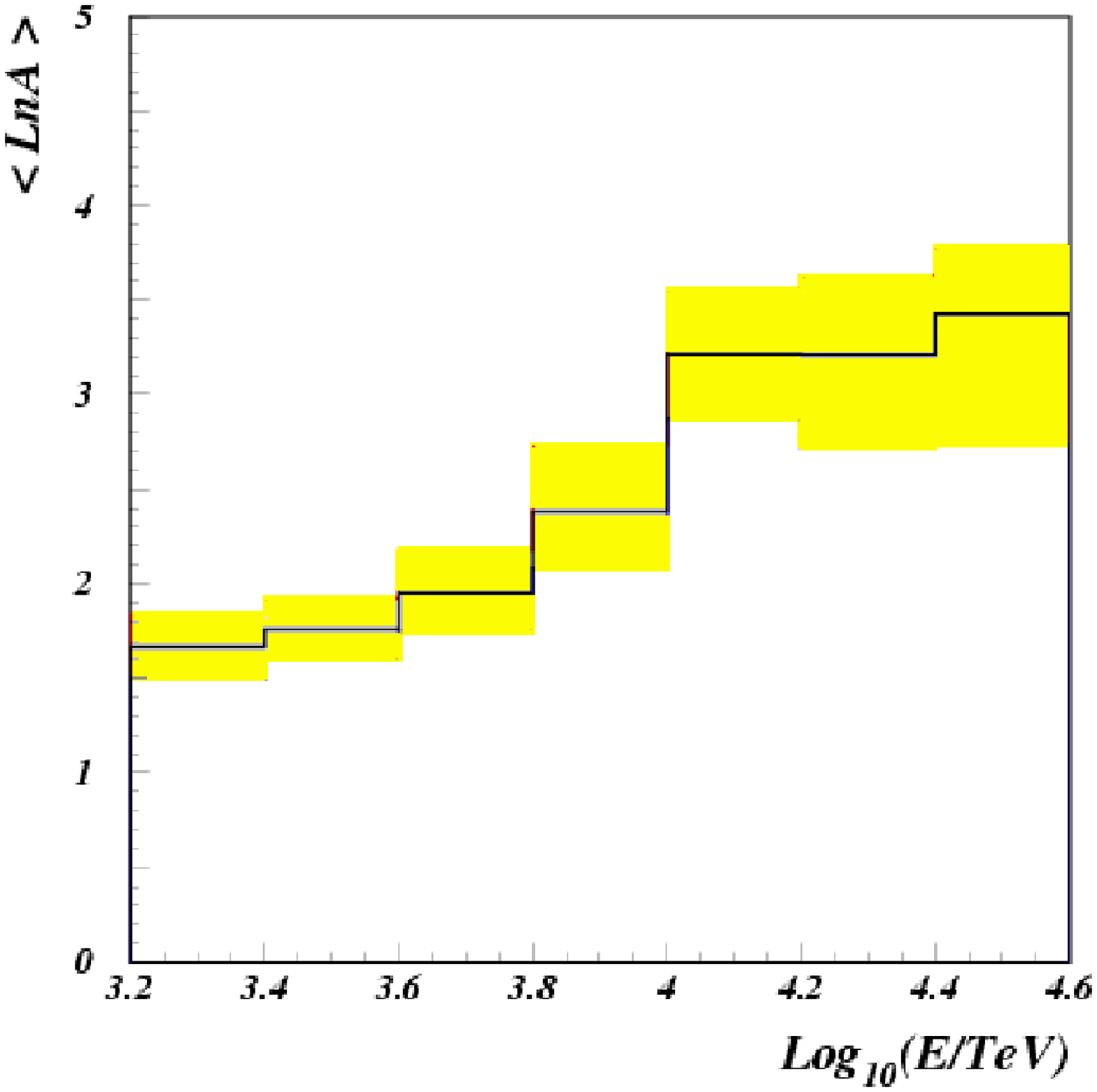} 
\epsfysize=8cm
         \epsffile{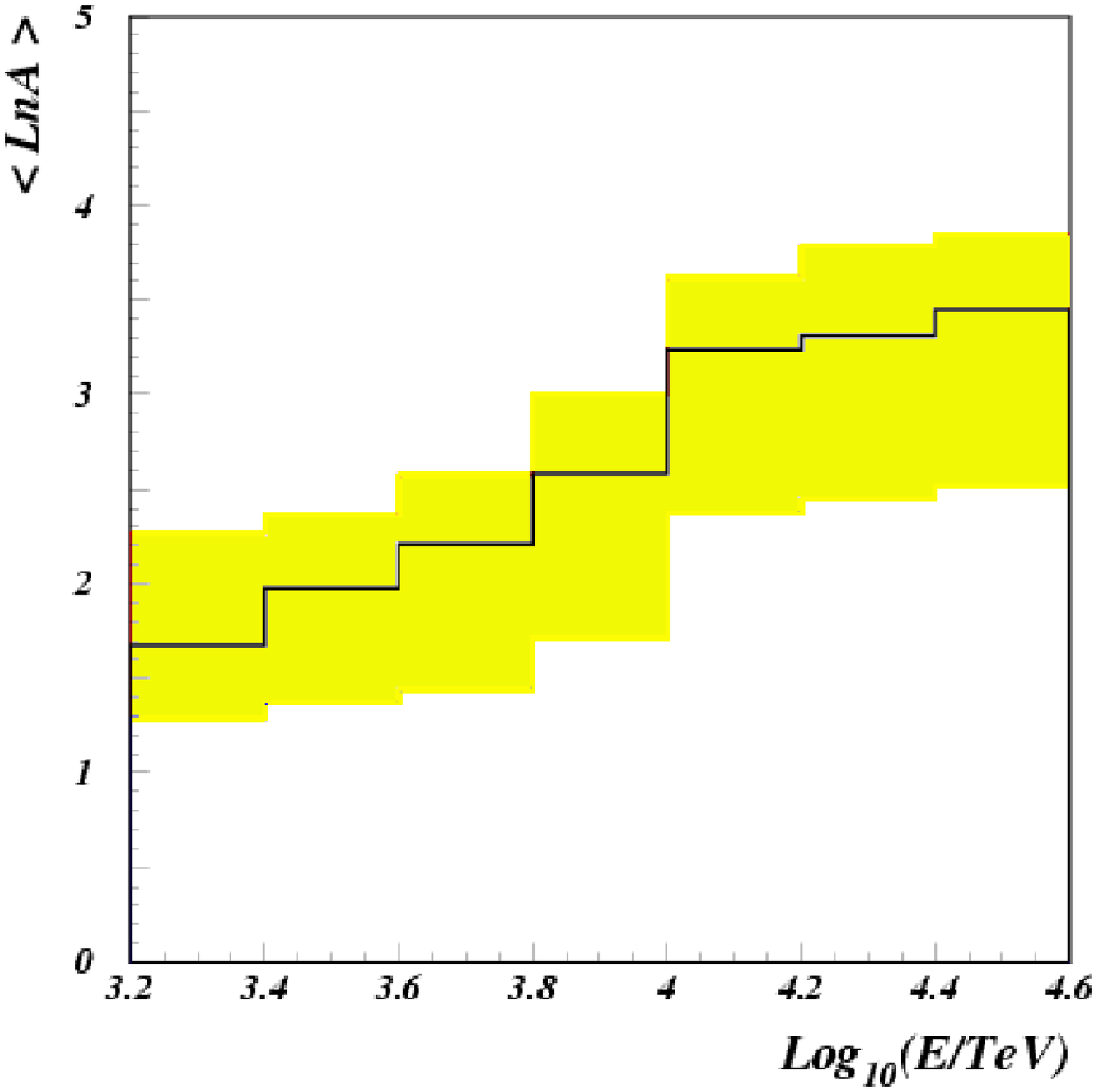} 
}   
{\small \hskip 5.0 truecm (a) \hskip 7.5 truecm (b)}
\caption{\label{fig21}\small EASTOP-MACRO coincidences. $\langle ln A
  \rangle$ vs primary energy for: (a) $p/Fe$  and (b) $Light/Heavy$ 
 compositions.
The histograms (black lines) are obtained from the data, the shaded areas include
 the   uncertainties discussed in the text.
}
\end{center}
\end{figure}

For each size bin we take from the simulation the $log_{10}(E)$ distributions of contributing mass groups weighted
by the parameters $p_1$ and $p_2$  with weights $w_k$ representing the
relative efficiency to trigger the underground apparatus.
The resulting distributions from different size bins are summed together,
and  we  obtain the simulated energy spectra of the two basic
components that reproduce the experimental data. 
The values of the fitting parameters $p_1$ and $p_2$ have been used to compute the average 
$\langle ln A \rangle$; Fig. \ref{fig21} shows $\langle lnA \rangle$ 
versus  $log_{10}E$ (E in TeV); the shaded regions  include the uncertainties in the fitting parameters  
$p_1$ and $p_2$ for (a) the p/Fe
composition model and (b) for the light/heavy model.

Fig. \ref{fig21} shows the results of the fits, plotted as  $\langle lnA \rangle$
versus  $log_{10}E$ (E in TeV); the shaded regions  include the
uncertainties in the fitting parameters  $p_1$ and $p_2$ for (a) the p/Fe
composition model and (b) for the light/heavy model.
The results show an increase of $\langle lnA \rangle$
with energy in the CR  knee region.
The results are in agreement with the measurements
of EAS-TOP alone at the surface using the same (QGSJET) interaction model.
Our data also agree with the results of the Kascade experiment.
\begin{figure}
 \vspace{-1.cm}
 \begin{center}
  \mbox{ \epsfysize=9.8cm
         \epsffile{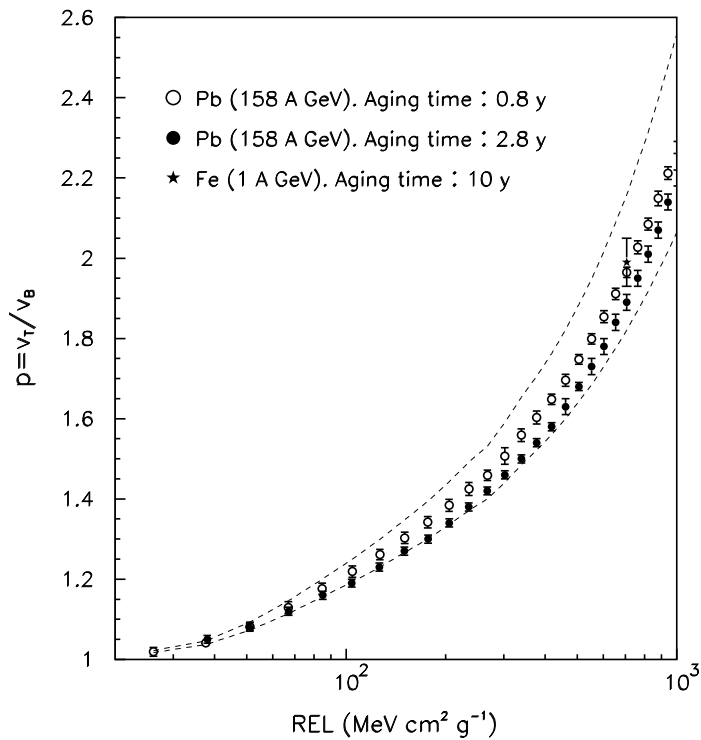} }
 \end{center}
\caption {\label{fig22}\small $p=v_T/v_B $ vs. REL for CR39 exposed to $^207 Pb^82+$ ions of 158 A GeV
and $^56Fe^26+$ ions of 1A GeV at different times after production. This was done to estimate possible
aging effects. The dashed lines indicate the systematic uncertainty arising mainly from fluctuations of
the bulk etching rate $v_B$.
}
\end{figure}
The EAS-TOP and MACRO coincidences offered the unique opportunity of 
measuring
the lateral distribution of Cherenkov light in the 10 $\div$ 100 TeV energy
range by associating the Cherenkov light collected by the EAS-TOP telescopes 
with 
the TeV muon  through  MACRO. 
We compared  the measured Cherenkov light lateral distribution with simulations based on the CORSIKA-QGSJET code used for the composition analysis;
this check provided an experimental validation of the code \cite{kna5}.

\section{Nuclear Track Detector Calibrations}

We performed further calibrations of the nuclear track detector CR39
with both slow and fast ions. In all measurements we have seen no
deviation of its response from the Restricted Energy Loss (REL) model.
To complete the calibration, nuclear track detector stacks of
CR\( 39 \) and Lexan foils, placed before and after various targets,
were exposed to 158 A \GeV \( \, \, Pb^{82+} \) ions at the CERN-SPS
and to 1 A \GeV \( \, \, Fe^{26+} \) ions at the BNL-AGS. In traversing
the target, the beam ions produce nuclear fragments with \( Z<82e \)
and \( Z<26e \) for the lead and iron beams, respectively; this allows
a measurement of the response of the detector in a \( Z \) region
relevant to the detection of magnetic monopoles. Previous analyses have shown that the CR39 charge resolution is about $0.19 e$ in the range 
$72 e \le Z \le 83e $ (obtained by measurements of the etch-cone
heights); at lower Z the measurement of the cone base diameters allow to separate the different charges. 
Tests were made looking for a possible dependence of the CR39 response
from its age, i.e. from the time elapsed between the date of production
and the date of exposure (``aging effect'' \cite{miriam}). Two sets
of sheets, 0.8 y and 2.5 y old,  were exposed in 
1994 to 158 A \GeV \( \, \, Pb^{82+} \) ions. For each detected nuclear
fragment the reduced etch rate \( p=v_{T}/v_{B} \) (\( v_{T} \)
and \( v_{B} \) are the track and bulk etching rates, respectively)
was computed and plotted in Fig. \ref{fig22} vs REL. The dashed lines represent
the systematic uncertainties coming mainly from the uncertainty on \( v_{B}
\). A recent test was made by 
exposing  10 years old CR39
samples to 1 A GeV  $Fe^{26+} $ ions; the detector response
is shown as a black star in Fig. \ref{fig22}. Thus  within experimental uncertainties, aging effects in the MACRO
CR39 are negligible.
Until now  we etched \( 821\, \, m^{2} \) of CR39 detectors, of which  \( 626\, \, m^{2} \)  have been completely analyzed. 
As no candidates were found, the CR39 90\% c.l. limit for an isotropic flux of monopoles with \( \beta >0.1 \) is  at the level of 
\( 2 \, \cdot 10^{-16}\, \, cm^{-2}s^{-1}sr^{-1} \).

\section{Search for Lightly Ionizing Particles}

Fractionally charged particles could be expected in Grand Unified Theories
as deconfined quarks; the expected charges range 
from Q=\textit{e}/5  to Q=\textit{e} 2/3. They should release a fraction $(Q/e)^2$
of the energy deposited by a muon traversing a medium. Lightly
Ionizing Particles (LIPs) have been searched for in MACRO using a four-fold
coincidence between three layers of scintillators and the streamer
tube system \cite{mac32}. The 90 \% c.l. flux upper limits for LIPs with
charges 2\textit{e}/3, 
\textit{e}/3  and \textit{e}/5 are presently at the level of 
$1.5 \cdot 10^{-15} cm^{-2}
s^{-1}sr^{-1}$.

\section{Conclusions}

The MACRO detector took data from 1989 to the end of year 2000. In 2001 we have extended most of our
analyses and searches. 
We would like to stress that MACRO obtained important results in all the items listed in the proposal :
\begin{itemize}
\item GUT Magnetic Monopoles. We now have the best flux upper limit over
the widest $\beta$ range, thanks to the large acceptance and the redundancy
of the different techniques employed. This limit value is a unique
result and it will stand for a long time.
\item Atmospheric neutrino oscillations. In this field MACRO has had its
major achievements. Analyses of different event topologies, different
energies, the exploitation of Coulomb multiple scattering in the detector
give strong support to the hypothesis of $\nu_\mu \rightarrow \nu_\tau$ oscillations.
\item High energy muon neutrino astronomy. MACRO has been highly competitive
with other underground experiments thanks to its good angular accuracy.
It has been limited only by its livetime and the size of the detector.
\item Search for bursts of $\bar{\nu}_e$
from stellar gravitational
collapses. In this field MACRO was sensitive to supernova events in the Galaxy,
it started  the SN WATCH  system, and for a certain time it was the only detector in operation.
\item Cosmic ray downgoing muons. MACRO  observed the shadows of primary cosmic rays by the Moon and the
Sun; this is also a proof of our pointing capability. We observed
the seasonal variation ($\sim 2\%$ amplitude) over many years. We observed solar
and sidereal variations 
with reasonable statistical
significances even if the amplitudes of the variations are small (
0.08\%). No excesses of secondary muons attributable to astrophysical
point sources (steady, modulated or bursting) were observed. The limits
obtained are the best of any underground detector.\\
We used multi-parameter fits and improved Monte
Carlo simulations to explore the CR composition around
the {}``knee{}'' of the primary CR energy spectrum.
\begin{figure}
 \vspace{-2.5cm}
  \begin{center}
  \mbox{
\epsfysize=14.8cm
         \epsffile{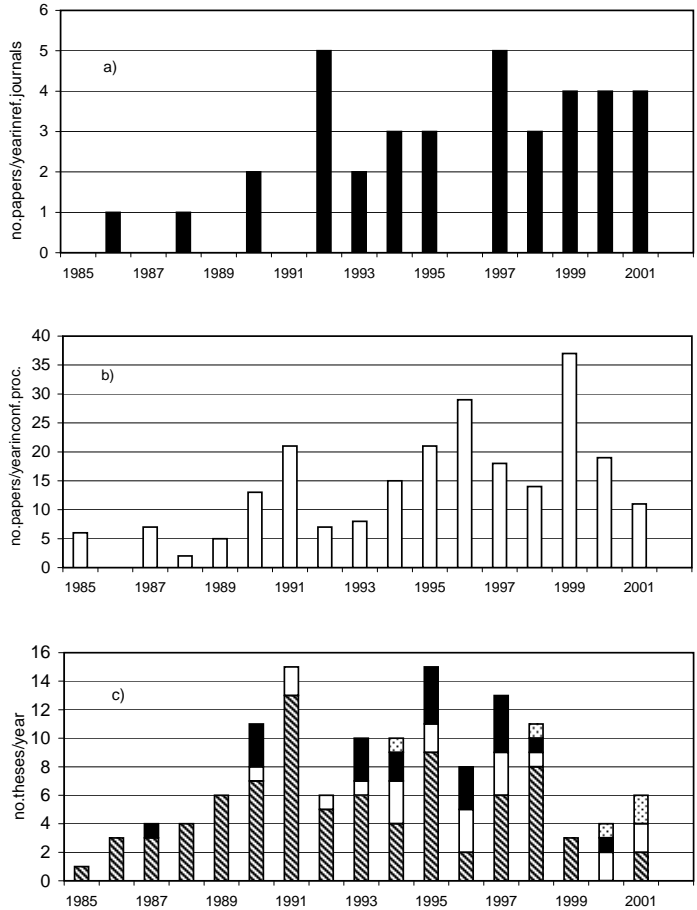}
}
 \vspace{-1.cm}
\caption{\label{fig23}\small Time evolution of the MACRO publications: (a)
  in refereed journals, (b) in proceedings of conferences, (c) MACRO
  theses (the dashed boxes indicate Laurea theses, black boxes American PhD
  theses, white boxes Italian PhD theses, points theses de Doctorat Nationales).}
\end{center}
\end{figure}
\item Results have been obtained by studying the coincidence events between
MACRO and the EASTOP array. This item represents a unique occasion
as no other two experiments in such configuration exist. The number
of events is limited due to the small common acceptance and short
combined livetime. The data indicate an increase with increasing energy
of the average Z of the primary CR nuclei.
\item Sensitive searches for exotic particles have been carried
out for possible Dark Matter  candidates : (i) WIMPs, looking
for upgoing muons  from the center of the Earth and of the Sun;
(ii) Nuclearites and Q-balls (obtained as byproducts of MM searches).
(iii) Other limits concern possible Lightly Ionizing Particles.
\end{itemize}

Several of the above results (in particular the multiple Coulomb scattering
analysis, the low energy neutrino data, etc) would have reached a
greater significance if MACRO could have been granted an extension
in data taking.

The dismantling of MACRO went regularly and essentially on schedule.
We recuperated part of the electronics (modules, circuits, cables,
etc) to be used in our Institutions, and donated the photomultipliers
and part of the streamer tubes to other experiments.

The MACRO scientific and technical results have been 

\begin{itemize}
\item published in 36 papers in refereed journals (we expect to publish
10 more with final results)
\item published in  226 contributions to  conferences and in invited papers
\item discussed in about 534 Internal Memos
\item used for 83 italian Laurea theses
\item used for 22 italian Dottorato theses
\item used for 23 US PhD theses
\item used for 5 moroccan theses of Doctorat Nationales
\end{itemize}

Fig. \ref{fig23} shows the time evolution of the published papers, conference proceedings
and of the theses (Laurea, Dottorato, PhD, Doctorat Nationale).

\subsection*{Acknowledgments}

{\small We acknowledge the support of the Directors and of the staff of the
Gran Sasso Laboratory and of the Institutions partecipating in the
experiment. We thank the Istituto Nazionale di Fisica Nucleare (INFN),
the US Department of Energy and the US National Science Foundation
for their support. We acknowledge the strong cooperation of our technical
staff, in particular of R. Assiro, E. Barbarito, A. Boiano, E. Bottazzi, P. Calligola, A. Candela,
A. Ceres, D. Cosson, P. Creti, L. Degli Esposti, U. Denni, D. Di Ferdinando,
R. Diotallevi, R. Farano, E. Favero, A. Frani,
M. Gebhard, R. Giuliani, M. Goretti, H. Gran, A. Hawthorne, A. Leone,
D. Margherita, V. Marrelli, F. Massera, S. Mengucci, M. Mongelli,
L. Mossbarger, M.Orsini, S. Parlati, G. Pellizzoni, M. Perchiazzi,
C. Pinto, A. Sacchetti, P. Saggese, S. Sondergaards, S. Stalio, M.
Vakili, C. Valieri and N. Zaccheo.
We thank INFN-FAI, ICTP (Trieste), NATO and WorldLab for providing
fellowships and grants for non-Italian citizens.


\begin{thebibliography}{M1}
\bibitem{mac1}{\small MACRO Collaboration, C. De Marzo et al., {}``MACRO: a
    large area detector at the Gran Sasso laboratory{}'', Nuovo Cimento 9C(1986)281.}
\bibitem{mac2}{\small MACRO Collaboration, M.Calicchio et al., {}``The
    MACRO detector at the Gran Sasso laboratory{}'', Nucl. Instr. Methods Phys. Res. A264(1988)18.}
\bibitem{mac3}{\small MACRO-EASTOP Collaborations, R.Bellotti et al.,
    M.Aglietta et al., {}``Simultaneous observation of the extensive air
    showers and deep underground muons at the Gran Sasso laboratory{}'', Phys. Rev. D42(1990)1396.}
\bibitem{mac4}{\small MACRO Collaboration, S.Ahlen et al., {}``Study of
    penetrating cosmic ray muons and search for large scale anisotropy at
    the Gran Sasso laboratory{}'', Phys. Lett. B249(1990)149.}
\bibitem{mac5}{\small MACRO Collaboration, S.Ahlen et al., {}``Arrival time
    distributions of very high energy cosmic ray muons in MACRO{}'',
    Nucl. Phys. B370(1992)432; LNGS-91/01.}
\bibitem{mac6}{\small MACRO Collaboration, S.Ahlen et al., {}``Study of the
    ultrahigh energy primary cosmic ray composition with the MACRO
    experiment{}'',  Phys. Rev. D46(1992)895.}
\bibitem{mac7}{\small MACRO Collaboration, S.Ahlen et al., {}``Measurement
    of the decoherence function with the MACRO detector at Gran Sasso{}'',
    Phys. Rev. D46(1992)4836; LNGS-92/29.}
\bibitem{mac8}{\small MACRO Collaboration, S.Ahlen et al., {}``Search for
    neutrino bursts  from collapsing stars with the MACRO detector{}'',
    Astroparticle Phys. 1(1992)11; LNGS-92/32.}
\bibitem{mac9}{\small MACRO Collaboration, S.Ahlen et al., {}``Search for
    nuclearites using the MACRO detector{}'', Phys. Rev. Lett. 69(1992)1860. }
\bibitem{mac10}{\small MACRO Collaboration, S.Ahlen et al., {}``First
    supermodule of the MACRO  detector at Gran Sasso{}'',
    Nucl. Inst. Meth. Phys. Res. A324(1993)337; LNGS 92/34.}
\bibitem{mac11}{\small MACRO Collaboration, S.Ahlen et al., {}``Muon
    astronomy with the MACRO  detector{}'', Astrophys. J. 412(1993)301.}
\bibitem{mac12}{\small MACRO Collaboration, S.Ahlen et al., {}``Search for
    slow moving magnetic  monopoles with the MACRO detector{}'', Phys. Rev. Lett. 72(1994)608; LNGS 93/84.}
\bibitem{mac13}{\small GRACE-MACRO Collaborations, M.Ambrosio et al.,
    {}``Coincident observation  of air Cherenkov light by a surface array
    and muon bundles by a  deep underground detector{}'', Phys. Rev. D50(1994)3046.}
\bibitem{mac14}{\small EASTOP-MACRO Collaborations, M.Aglietta et al.,
    {}``Study of the primary cosmic  ray composition around the knee of the
    energy spectrum{}'', Phys. Lett. B337(1994)376.}
\bibitem{mac15}{\small MACRO Collaboration, M.Ambrosio et al.,
    {}``Performance of the MACRO streamer tube system in the search for
    magnetic monopoles{}'', Astroparticle Phys. 4(1995)33; LNGS 95/11.}
\bibitem{mac16}{\small MACRO Collaboration, M.Ambrosio et al., {}``Vertical
    muon intensity measured with MACRO at the Gran Sasso laboratory{}'',
    Phys. Rev. D52(1995)3793.}
\bibitem{mac17}{\small MACRO Collaboration, M.Ambrosio et al.,
    {}``Atmospheric neutrino flux measurements using upgoing muons{}'',
    Phys. Lett. B357(1995)481.} 
\bibitem{mac18}{\small MACRO Collaboration, M.Ambrosio et al., {}``The
    performance of MACRO liquid scintillator in the search for magnetic
    monopoles with $10^{-3} \le \beta \le  1 $'', Astroparticle Phys. 6(1997)113; INFN-AE 96/22.}
\bibitem{mac19}{\small MACRO Collaboration, M.Ambrosio et al. {}``High
    energy cosmic ray physics with the MACRO detector at Gran Sasso: Part
    I. Analysis methods and experimental results{}'', Phys. Rev. D56(1997)1407; INFN-AE 96/28.}
\bibitem{mac20}{\small MACRO Collaboration, M.Ambrosio et al., {}``High
    energy cosmic ray physics with the MACRO detector at Gran Sasso: Part
    II.  Primary spectra and composition{}'', Phys. Rev. D56(1997)1418; INFN-AE 96/29.}
\bibitem{mac21}{\small MACRO Collaboration, M.Ambrosio et al., {}``Seasonal
    variations in the underground muon intensity as seen by MACRO{}'',
    Astroparticle Phys. 7(1997)109; INFN-AE 97/05. }
\bibitem{mac22}{\small MACRO Collaboration, M.Ambrosio et al., {}``Magnetic
    monopole search with the MACRO detector at Gran Sasso{}'',
    Phys. Lett. B406(1997)249;  INFN-AE 97/16.}
\bibitem{mac23}{\small MACRO Collaboration, M.Ambrosio et al. {}``Real time
    supernova neutrino burst detection with MACRO{}'', Astroparticle
    Phys. 8(1998)123;  INFN-AE 97/44.}
\bibitem{mac24}{\small MACRO Collaboration, M.Ambrosio et al., {}``The
    observation of upgoing charged particles produced by high energy muons
    in underground detectors{}'', Astroparticle Phys. 9(1998)105; INFN-AE 97/55; hep-ex/9807032.}
\bibitem{mac25}{\small MACRO Collaboration, M.Ambrosio et
    al. {}``Measurement of the atmospheric neutrino-induced upgoing muon
    flux using MACRO{}'', Phys. Lett. B434(1998)451; INFN-AE 98/13; hep-ex/9807005.}
\bibitem{mac26}{\small MACRO Collaboration, M.Ambrosio et al.,
    {}``Observation of the shadowing  of cosmic rays by the Moon using a
    deep underground detector{}'',  Phys. Rev. D 59(1999)012003; INFN-AE 98/14; hep-ex/9807006.}
\bibitem{mac27}{\small MACRO Collaboration, M.Ambrosio et al.,
    {}``Measurement of the energy spectrum of underground muons at Gran
    Sasso with a transition radiation detector{}'', Astroparticle
    Phys. 10(1999)11; INFN-AE 98/15; hep-ex/9807009.}
\bibitem{mac28}{\small MACRO Collaboration, M.Ambrosio et al., {}``Limits
    on dark matter WIMPs using upward-going muons in the MACRO
    detector{}'', Phys. Rev. D60(1999)082002; hep-ex/9812020.}
\bibitem{mac29}{\small MACRO Collaboration, M.Ambrosio et al., {}``High
    statistics measurement of  the underground muon pair separation at Gran
    Sasso{}'', Phys. Rev.  D60(1999)032001; hep-ex/9901027; INFN-AE 99/04 (1999).}
\bibitem{mac30}{\small MACRO Collaboration, M.Ambrosio et al.,
    {}``Nuclearite search with the MACRO  detector at Gran Sasso{}'',
    Eur. Phys. J. C13(2000)453; hep-ex/9904031. }
\bibitem{mac31}{\small MACRO Collaboration, M.Ambrosio et al., {}``Low
    energy atmospheric muon neutrinos in MACRO{}'', Phys. Lett. B478(2000)5; hep-ex/0001044.}
\bibitem{mac32}{\small MACRO Collaboration, M. Ambrosio et al., {}``A
    search for lightly ionizing particles with the MACRO detector{}'',
    Phys. Rev. D62(2000)052003; hep-ex/0002029. }
\bibitem{mac33}{\small MACRO Collaboration, M. Ambrosio et al.,
    {}``Neutrino astronomy with the MACRO detector{}''
    Astrophys. J. 546(2001)1038; astro-ph/0002492.}
\bibitem{mac34}{\small MACRO Collaboration, M.Ambrosio et al., {}``Matter
    effects in upward-going muons and sterile neutrino oscillations{}'',
    Phys. Lett. B517(2001)59;  hep-ex/0106049.}
\bibitem{mac35}{\small MACRO Collaboration, M.Ambrosio et al., {}``The
    MACRO detector at Gran Sasso{}'', accepted for publication on NIM A. }
\bibitem{mac36}{\small MACRO Collaboration, M.Ambrosio et al., {}``A
    combined analysis technique  for the search for fast magnetic monopoles
    with the MACRO detector{}'', hep-ex/0110083; accepted for publication on Astroparticle Physics. }
\bibitem{mac37}{\small P. Bernardini, {}``Neutrino astronomy using
    upward-travelling muons in MACRO{}'', Invited talk at {}``Very High
    Energy Phenomena in the Universe{}'', XXXVI Rencontres de Moriond, Les Arcs (2001).}
\bibitem{mac52}{\small D. Bakari for the MACRO Collaboration, {}``Estimate
    of the energy of upgoing muons with multiple Coulomb scattering{}'',
    hep-ex/0105087, Proceedings of the NATO ARW on Cosmic Radiations: from
    Astronomy to Particle Physics,   Oujda (Morocco), 21-23 March 2001.}
\bibitem{mac52_a}{\small G. Battistoni for the MACRO Collaboration,
    {}``Neutrino induced upgoing muon  energy estimation by multiple
    scattering with MACRO{}'',  Proceedings of the NATO ARW on Cosmic
    Radiations: from Astronomy to Particle  Physics,  Oujda (Morocco), 21-23 March 2001.}
\bibitem{mac56}{\small M. Cozzi and L. Patrizii for the MACRO
    Collaboration, {}``Nuclear track detectors.  Searches for magnetic
    monopoles and for nuclearites{}'',  Proceedings of the NATO ARW on
    Cosmic Radiations: from  Astronomy to Particle Physics,  Oujda (Morocco), 21-23 March 2001. }
\bibitem{mac52_b}{\small H. Dekhissi for the MACRO Collaboration,
    {}``Muon astronomy with the underground detectors{}'', Proceedings of
    the  NATO ARW on Cosmic Radiations: from Astronomy to Particle Physics,
    Oujda (Morocco),  21-23 March 2001.}
\bibitem{mac53}{\small M. Grassi for the MACRO Collaboration, {}``A search
    for gravitational stellar collapses{}'', Proceedings of the NATO ARW on
    Cosmic Radiations: from  Astronomy to Particle Physics,  Oujda (Morocco), 21-23 March 2001.}
\bibitem{mac38}{\small S. Kyriazopoulou for the MACRO Collaboration,
    {}``Search for slow magnetic monopoles with the MACRO scintillation
    detector{}'', Proceedings of the  NATO ARW on Cosmic Radiations: from
    Astronomy to Particle Physics, Oujda (Morocco), 21-23 March 2001.}
\bibitem{mac38_a}{\small F. Maaroufi and A. Margiotta for the MACRO
    Collaboration, {}``Daily variation studies with the MACRO detector{}'',
    Proceedings of the NATO ARW on  Cosmic Radiations: from Astronomy to Particle Physics,  Oujda
    (Morocco), 21-23 March 2001.}
\bibitem{mac38_b}{\small T. Montaruli for the MACRO Collaboration, {}``High
    energy neutrino astronomy and WIMP search results{}'', Proceedings of
    the  NATO ARW on Cosmic Radiations: from Astronomy to Particle Physics,  Oujda
    (Morocco), 21-23 March 2001.}
\bibitem{mac38_c}{\small L. Perrone for the MACRO Collaboration, {}``Search
    for a diffuse neutrino  flux from astrophysical sources{}'',
    Proceedings of the NATO ARW on Cosmic Radiations: from Astronomy to
    Particle Physics,  Oujda  (Morocco), 21-23 March 2001. }
\bibitem{mac39}{\small F. Ronga for the MACRO  Collaboration,
    {}``Atmospheric neutrinos and neutrino oscillations in the MACRO
    experiment{}'',   Proceedings of the NATO ARW on   Cosmic Radiations:
    from Astronomy to  Particle Physics,  Oujda (Morocco), 21-23 March 2001. }
\bibitem{mac55}{\small M. Sitta for the MACRO Collaboration, {}``Monopole
    catalysis of nucleon decay: theory and experimental results{}'',
    Proceedings of the NATO ARW on Cosmic Radiations: from Astronomy to
    Particle Physics,   Oujda (Morocco), 21-23 March 2001. }
\bibitem{mac40}{\small I. De Mitri for the MACRO Collaboration, {}``Search
    for magnetic monopoles in the cosmic radiation with the MACRO detector
    at Gran Sasso {}``, International Europhysics Conference on High Energy
    Physics,  Budapest,  July 12-18 2001. }
\bibitem{mac41}{\small E. Scapparone et al. for the MACRO Collaboration,
    {}``Study  of neutrino induced upgoing muon energy{}'', International
    Europhysics Conference on  High Energy Physics, Budapest, July 12-18 2001.}
\bibitem{mac41a}{\small G. Giacomelli and M. Giorgini for the MACRO
    Collaboration, {}``Atmospheric neutrino oscillations in MACRO{}'',
    Invited talk at NO-VE, Intern. Workshop on Neutrino Oscillation in
    Venice, 24-26 July 2001}
\bibitem{mac57}{\small F. Cei for the MACRO Collaboration, {}``Search for
    lightly ionizing particles with the MACRO detector{}'', ICRC 2001,
    Hamburg, Germany, 7-15 August 2001.}
\bibitem{mac45}{\small N. Giglietto for the MACRO Collaboration, {}``Moon
    and Sun shadowing effect on the MACRO apparatus{}'', ICRC 2001,
    Hamburg, Germany, 7-15 August 2001.}
\bibitem{mac46}{\small T. Montaruli for the MACRO Collaboration, {}``Final
    results on atmospheric neutrino oscillations with MACRO{}'', ICRC 2001,
    Hamburg, Germany 7-15 August 2001. }
\bibitem{mac42}{\small S. Mufson for the MACRO Collaboration,
    {}``Measurement of the Solar Diurnal and Sidereal Muon Waves with
    MACRO{}'', ICRC 2001, Hamburg, Germany, 7-15 August 2001.}
\bibitem{mac48}{\small L. Perrone for the MACRO Collaboration, {}``Neutrino
    astronomy with MACRO{}'', ICRC 2001, Hamburg, Germany, 7-15 August 2001.}
\bibitem{mac44}{\small M. Sioli for the MACRO Collaboration,
    {}``Measurement of the ratio double/single muon events as a function of rock depth with MACRO{}'',
    hep-ex/0201017;{}``Use of Coulomb scattering for determining neutrino energies with MACRO{}'',
    hep-ex/0201016, ICRC 2001, Hamburg, Germany, 7-15 August 2001. }
\bibitem{mac47}{\small M. Sitta for the MACRO Collaboration, {}``Rare
    particle searches with MACRO{}'', ICRC 2001, Hamburg, Germany, 7-15 August 2001. }
\bibitem{mac54}{\small M. Spurio for the MACRO Collaboration, {}`` Low
    energy atmospheric $\nu_\mu$ measurements{}'', ICRC 2001, Hamburg, Germany, 7-15 August 2001. }
\bibitem{mac43}{\small P. Vallania for the EASTOP-MACRO Collaboration,
    {}``The primary CR composition around the knee from EAS e.m., GeV and
    TeV muon data{}'', ICRC 2001, Hamburg, Germany, 7-15 August 2001.}
\bibitem{mac48a}{\small G. Battistoni for the MACRO Collaboration,
    {}``Cosmic ray composition around the knee from EAS electromagnetic and
    muon data{}'', astro-ph/0112473; {}``Study of cosmic ray primaries  and their cascade at
    $E_o = 10 - 100 $ \TeV through EAS-TOP and MACRO{}'', astro-ph/0112475;
    TAUP 2001, Laboratori Nazionali del Gran Sasso, September 8-12 2001.}
\bibitem{mac49}{\small I. De Mitri for the MACRO Collaboration, {}``Search
    for massive rare particles in MACRO{}'', TAUP 2001, Laboratori
    Nazionali del Gran Sasso, September 8-12 2001.}
\bibitem{mac50}{\small E. Scapparone for the MACRO Collaboration, {}``Study
    of neutrino induced upgoing muon energy with MACRO{}'', TAUP 2001, LNGS, September 8-12, 2001.}
\bibitem{mac51}{\small M. Giorgini for the MACRO Collaboration,
    {}``Performance of the MACRO limited streamer tubes for estimates of
    muon energies {}'', 7th Int. Conf. on Advanced Technology and Particle
    Physics, Como, Italy (2001).}
\bibitem{bartol}{\small V. Agrawal et al., Phys. Rev. D53(1996)1314.} 
\bibitem{gluck}{\small M. Gluck et al.,  Z. Phys.  C67(1995)433.  }
\bibitem{lohmann}{\small W. Lohmann et al., ``Energy loss of muons in the
    energy range 1 - 10000 GeV''  CERN 85-03.} 
\bibitem{feldman}{\small G. Feldman and R. Cousins, Phys. Rev. D57(1998)3873. }
\bibitem{mactesi}{\small M. Giorgini, {}``Study of atmospheric neutrino
    oscillations by energy estimates of upgoing muons in MACRO{}'', Tesi di
    Dottorato, Universit\`{a} di  Bologna (2002). }
\bibitem{sk}{\small SuperKamiokande Coll., Y.Fukuda et al.,
    Phys. Rev. Lett. 81(1998)1562; Phys. Lett.   B433(1998)9;
    Phys. Rev. Lett. 85(2000)3999;  Nucl. Phys. B Proc. Suppl. 91(2001)127; T. Toshito, hep-ex/0105023 (2001).}
\bibitem{soud2}{\small Soudan 2 Coll.,  W.W.M. Allison et al., Phys. Lett.
    B391 (1997) 491;  Phys. Lett.  B449 (1999) 137; W. Anthony Mann,
    hep-ex/0007031 (2000); T. Mann et al., Nucl. Phys. B Proc. Suppl. 91 (2001) 134.}
\bibitem{lipari94}{\small P. Lipari et al., Phys. Rev. Lett. 74(1995)384. }
\bibitem{bottino}{\small A. Bottino, N. Fornengo, F. Donato and S. Scopel
    (private communication). N. Fornengo, in Proceedings of the Ringberg
    Euroconference ``New Trends in Neutrino Physics'', Ringberg Castle,
    Tegernsee, Germany,  1998, edited  by B. Kniel, World Scientific, Singapore. }
\bibitem{dama}{\small R. Bernabei et al., Phys. Lett. B389(1996)757. }
\bibitem{giapat}{\small G. Giacomelli and L. Patrizii, ``Magnetic
    monopoles'', Lecture at the Fifth School on Particle Astrophysics,
    Trieste  29 June-10 July 1998, hep-ex/0002032. \\ G. Giacomelli et
    al. (Magnetic Monopole Bibliography)  hep-ex/0005041.}
\bibitem{derka}{\small J. Derkaoui et al., Astropart. Phys. 9(1998)173; Astropart. Phys. 10(1999)339.}
\bibitem{nucl1}{\small S. Nakamura et al., Phys. Lett. B263(1991)529.  }
\bibitem{nucl2}{\small S. Orito et al., Phys. Rev. Lett. 66(1991)1951.  }
\bibitem{bakari}{\small D. Bakari et al., {}``Magnetic monopoles,
    nuclearites, Q-balls: a qualitative  picture{}'', hep-ex/0004019. }
\bibitem{nucl3}{\small P. B. Price, Phys. Rev.  D38(1988) 3813. }
\bibitem{nucl4}{\small D. Ghosh and S. Chatterjea, Europhys. Lett. 12(1990)25. }
\bibitem{tesiphd}{\small M. Sioli, {}``A new approach to the study of high
    energy muon bundles with the MACRO detector at Gran Sasso{}'', Tesi di
    Dottorato, Universit\`{a} di  Bologna (2000). }
\bibitem{corint}{\small P. Lipa, P. Carruthers, H.C. Eggers and B. Buschbeck, Phys. Lett. 285B(1992)300. }
\bibitem{clusters}{\small G. Battistoni et al., LNGS-95-09 (1995);
    Proceedings of the XXIV Int. Cosmic Ray Conf., Roma,  1995,
    ed. N. Iucci et al., Arti Grafiche Editoriali, Urbino, (1995), Vol. 1, p. 508. }
\bibitem{elbert}{\small J. W. Elbert et al, XVII ICRC (Paris, 1981) Vol. 7, p.42.}
\bibitem{ryazhaskaya}{\small O. G. Ryazhskaya for the LVD Collaboration,
    Nucl. Phys. (Proc. Suppl.) B87(2000)423.}
\bibitem{antonioli}{\small P. Antonioli et al., Astrop. Phys. 7(1997)357.  }
\bibitem{capde}{\small J.N. Capdevielle et al., KFK Report (1992)4998; FZKA (1998)6019. }
\bibitem{ref1}{\small EAS-TOP Collaboration, M. Aglietta et al., Nucl. Phys. B54B(1997)263. }
\bibitem{kna5}{\small J. Knapp and D. Heck, Extensive Air Shower Simulation with CORSIKA 5.61 (1998).}
\bibitem{as2}{\small M. Aglietta  et al., Nucl. Instr. \& Meth. A336(1993)310.}
\bibitem{as5}{\small EAS-TOP Coll., Astroparticle Physics, 10, 1, 1999 and Proc. 26th ICRC, 1, 230, (1999).}
\bibitem{miriam}{\small S. Cecchini et al., ``New calibrations and time
    stability of the INTERCAST CR-39'', Radiat. Meas. 34(2001)55, hep-ex/0104022. }
\end{thebibliography}
\end{document}